\documentclass[preprint]{elsarticle}

\usepackage{psfig}
\usepackage{graphicx}
\usepackage{pstricks}

\usepackage{amsmath}
\usepackage{amsfonts}
\usepackage{amssymb}
\usepackage{verbatim}
\usepackage{thanks}
\usepackage{latexsym}
\usepackage{enumerate}

\usepackage{graphicx}        

\newcommand{\NN}{\mathbb{N}}  

\newcommand{\IF}{\mbox{{\bf if}\ }}
\newcommand{\FI}{\mbox{{\bf fi}}}
\newcommand{\DO}{\mbox{{\bf do}\ }}
\newcommand{\OD}{\mbox{{\bf od}}}
\newcommand{\WHILE}{\mbox{{\bf while}\ }}
\newcommand{\END}{\mbox{{\bf end}}}
\newcommand{\THEN}{\mbox{\ {\bf then}\ }}

\newcommand{\ELSE}{\mbox{\ {\bf else}\ }}

\newcommand{\T}{\mbox{{\bf true}}}
\newcommand{\F}{\mbox{{\bf false}}}
\newcommand{\FAIL}{{\bf fail}}
\newcommand{\ES}{\mbox{$\emptyset$}}

\newcommand{\ra}{\mbox{$\:\rightarrow\:$}}
\newcommand{\tra}{\mbox{$\:\rightarrow^*\:$}}

\newcommand{\A}{\mbox{$\ \wedge\ $}}

\newcommand{\Or}{\mbox{$\ \vee\ $}}
\newcommand{\U}{\mbox{$\:\cup\:$}}
\newcommand{\myI}{\mbox{$\:\cap\:$}}  
\newcommand{\sse}{\mbox{$\:\subseteq\:$}}

\newcommand{\Mo}{\mbox{$\:\models\ $}}

\newcommand{\Msp}{\mbox{$\:\models_{\it sp}\ $}}

\newcommand{\te}{\mbox{$\exists$}}

\newcommand{\LL}{\mbox{$\ldots$}}

\newcommand{\MS}[1]{\mbox{${\cal M}[\![{#1}]\!]$}}

\newcommand{\MSP}[1]{\mbox{${\cal M}_{{\it sp}}[\![{#1}]\!]$}}

\newcommand{\NS}[1]{\mbox{${\cal N}[\![{#1}]\!]$}}

\newcommand{\ITE}[3]{\mbox{$\IF {#1} \THEN {#2} \ELSE {#3}\ \FI$}}
\newcommand{\IT}[2]{\mbox{$\IF {#1} \THEN {#2}\ \FI$}}

\newcommand{\WDD}[2]{\mbox{$\WHILE {#1}\ \DO {#2}\ \OD$}}

\newcommand{\HT}[3]{\mbox{$\{{#1}\}\ {#2}\ \{{#3}\}$}}

\newcommand{\sg}{{\mbox{$\sigma$}}}

\newcommand{\B}[1]{\mbox{$[\![{#1}]\!]$}}       
\newcommand{\C}[1]{\mbox{$\{{#1}\}$}}         

\newcommand{\NI}{\noindent} 

\newcommand{\HB}{\qed \III}

\newcommand{\VV}{\vspace{5 mm}}
\newcommand{\III}{\vspace{3 mm}}

\newcommand{\ignore}[1]{}  

\def\mylabel{\label}

\def\myref{\ref}

\def\nlni{\par\ifvmode\removelastskip\fi\vskip\baselineskip\noindent}

\newcommand{\this}{{\bf this}}
\newcommand{\Object}{{\bf object}}

\newcommand{\nulll}{{\bf null}}
\newcommand{\new}{{\bf new}}

\newcommand{\Objects}{{\cal D}_\Object}

\newcommand{\BEGIN}{\mbox{{\bf begin}}}
\newcommand{\block}[1]{\mbox{$\BEGIN \ {#1}\ \END$}}
\newcommand{\local}{\mbox{{\bf local}\ }}

\newtheorem{theorem}{Theorem}[section]
\newtheorem{defined}[theorem]{Definition}
\newenvironment{definition}{\begin{defined} \rm}{\end{defined}}
\newtheorem{exa}[theorem]{Example}
\newenvironment{example}{\begin{exa} \rm}{\end{exa}}

\newtheorem{lemma}[theorem]{Lemma}
\newtheorem{corollary}[theorem]{Corollary}

\newtheorem{note}[theorem]{Note}

\newenvironment{Theorem}{\begin{theorem}}{\end{theorem}}
\newenvironment{Example}{\begin{example}}{\end{example}}
\newenvironment{Lemma}{\begin{lemma}}{\end{lemma}}
\newenvironment{Def}{\begin{definition}}{\end{definition}}
\newenvironment{Note}{\begin{note}}{\end{note}}
\newenvironment{Cor}{\begin{corollary}}{\end{corollary}}

\newenvironment{Theorem-HB}{\begin{theorem}}{\HB\end{theorem}}
\newenvironment{Example-HB}{\begin{Example}}{\HB\end{Example}}
\newenvironment{Lemma-HB}{\begin{Lemma}}{\HB\end{Lemma}}
\newenvironment{Def-HB}{\begin{Def}}{\HB\end{Def}}
\newenvironment{Note-HB}{\begin{Note}}{\HB\end{Note}}
\newenvironment{Cor-HB}{\begin{Cor}}{\HB\end{Cor}}
\newenvironment{Warn-HB}{\begin{Warn}}{\HB\end{Warn}}

\newcommand{\mynewrule}[4]{
   \newenvironment{#1}{\nlni\begingroup
                       \refstepcounter{#3}
                       {\bf #2 \csname the#3\endcsname#4}
                      }
                      {\endgroup\vskip\baselineskip}
   }
\newcounter{RuleCnt}
\mynewrule{Rule}{RULE}{RuleCnt}{:}
\mynewrule{Axiom}{AXIOM}{RuleCnt}{:}

\newcounter{AuxRuleCnt}
\mynewrule{AuxRule}{RULE}{AuxRuleCnt}{:}
\mynewrule{AuxAxiom}{AXIOM}{AuxRuleCnt}{:}

\newcommand{\mytheoremname}[1]{{\bf (#1)}}

\newenvironment{Proof}
      {\medskip\noindent{\bf Proof.}}
      {\hfill$\Box$\medskip}

\begin{document}

\author[cwi,uva]{Krzysztof R. Apt\corref{cor1}} 
\ead{apt@cwi.nl}
\author[cwi,leiden]{Frank S. de Boer}
\ead{F.S.de.Boer@cwi.nl}
\author[oldenburg]{Ernst-R\"{u}diger Olderog}
\ead{olderog@informatik.uni-oldenburg.de}
\author[cwi,leiden]{Stijn de Gouw}
\cortext[cor1]{Corresponding author}

\address[cwi]{Centre for Mathematics and Computer Science (CWI), Amsterdam, The Netherlands} 
\address[uva]{University of Amsterdam, Institute of Language, Logic and Computation, Amsterdam}
\address[leiden]{Leiden Institute of Advanced Computer Science, University of Leiden, The Netherlands}
\address[oldenburg]{Department of Computing Science, University of Oldenburg, Germany}

\title{Verification of Object-Oriented Programs: \\
a Transformational Approach}


\begin{abstract}
  We show that verification of object-oriented programs by means of
  the assertional method can be achieved in a simple way by exploiting
  a syntax-directed transformation from object-oriented programs to
  recursive programs.  This transformation suggests natural proofs
  rules and its correctness helps us to establish soundness and
  relative completeness of the proposed proof system.  One of the
  difficulties is how to properly deal in the assertion language with
  the instance variables and aliasing.  The discussed programming
  language supports arrays, instance variables, failures and recursive
  methods with parameters.  We also explain how the transformational
  approach can be extended to deal with other features of
  object-oriented programming, like classes, inheritance,
  subtyping and dynamic binding.
\end{abstract}

\maketitle

\section{Introduction}
\label{sec:intro}

\subsection{Background and motivation}

Ever since its introduction in \cite{Hoa69} the assertional method has
been one of the main approaches to program verification.  Initially
proposed for the modest class of \textbf{while} programs, it has been
extended to several more realistic classes of programs, including
recursive programs (starting with \cite{Hoa71-proc}), programs with
nested procedure declarations (see \cite{LO80}), parallel programs
(starting with \cite{OG76a} and \cite{OG76b}), and distributed
programs based on synchronous communication (see \cite{AFR80}). At the
same time research on the theoretical underpinnings of the proposed
proof systems resulted in the introduction in \cite{Coo78} of the
notion of relative completeness and in the identification of the inherent
incompleteness for a comprehensive ALGOL-like programming language (see
\cite{Cla79}).

However, (relative) completeness of proof systems proposed for current
object-oriented programming languages (see the related work section
below) remained largely beyond reach because of the many intricate
and complex features of languages like Java.  In this paper we present
a transformational approach to the formal justification of proof
systems for object-oriented programming languages.  We focus on the
following main characteristics of objects:
\begin{itemize}
\item
objects possess (and {\em encapsulate}) their own (so-called instance) variables, and
\item
objects interact via {\em method} calls.
\end{itemize}
The execution of a method call  involves a
temporary {\em transfer} of control from the local state of the caller object to
that of the called object (also referred to by {\em callee}).  
Upon termination of the method call the control returns to the local state of the caller.
The method calls are the \emph{only way} to transfer control from 
one object to another.
We illustrate our approach by a syntax-directed transformation of
the considered object-oriented programs to \emph{recursive programs}. This transformation
naturally suggests the corresponding proof rules.
The main result of this paper is that the transformation  preserves  (relative) completeness.

To make this approach work a number of subtleties need to be
taken care of.  To start with, the `base' language needs to be
appropriately chosen. More precisely, to properly deal with the
problem of avoiding methods calls on the $\nulll$ object we need a
failure statement. In turn, to deal in a simple way with the
call-by-value parameter mechanism we use parallel assignment and block
statement.  Further, to take care of the local variables of objects at the
level of assertions we need to appropriately define the assertion
language and deal with the substitution and aliasing.

We introduced this approach to the verification of object-oriented
programs in our recent book \cite{ABO09} where we proved soundness.
The aim of this paper is to provide a systematic and self-contained
presentation which focuses on (relative) completeness and to explain
how to extend this approach to other features of object-oriented
programming.  Readers interested in example correctness proofs may
consult \cite[pages 226--237]{ABO09}.

\subsection{Related work}

The origins of the proof theory for recursive method calls presented
here can be traced back to \cite{Boer91}. However, in \cite{Boer91}
the transformational approach to soundness and relational completeness
was absent and failures were not dealt with.  In \cite{PdeB05} an
extension to the typical object-oriented features of {\em inheritance}
and {\em subtyping} is described.  There is a large literature on
assertional proof methods for object-oriented languages, notably for
Java.  For example, \cite{Jacobs04} discusses a weakest precondition
calculus for Java programs with annotations in the Java Modeling
Language (JML). JML can be used to specify Java classes and interfaces
by adding annotations to Java source files.  An overview of its tools
and applications is provided in \cite{Burdy05}.  In \cite{HJ00} a
Hoare logic for Java with abnormal termination caused by failures is
described.  However, this logic involves a major extension of the
traditional Hoare logic to deal with failures for which the
transformational approach breaks down.

Object-oriented programs in general give rise to dynamically evolving
{\em pointer} structures as they occur in programming languages like
Pascal.  This leads to the problem of \emph{aliasing}.  There is a
large literature on logics dealing with aliasing.
One of the early approaches, focusing on the linked data structures,
is described in \cite{M82}.  A more recent approach is that of {\em
  separation logic} described in \cite{Reynolds02}.  In \cite{Al03} a
Hoare logic for object-oriented programs is introduced based on an
explicit representation of the global store in the assertion language.
In \cite{BN05} restrictions on aliasing are introduced to ensure
encapsulation of classes in an object-oriented programming language
with pointers and subtyping.

Recent work on assertional methods for object-oriented programming
languages (see for example \cite{BarnettCDJL05}) focuses on {\em
  object invariants} and a corresponding methodology for {\em modular}
verification.  In \cite{MPL06} also a class of invariants is
introduced which support modular reasoning about complex object
structures.

Formal justification of proof systems for object-oriented programming languages
have been  restricted to soundness (see for example \cite{oheimb01} and \cite{klein06}).
Because of the many intricate and complex features of current object-oriented
programming languages
(relative) completeness remained largely beyond reach.
Interestingly, in the above-mentioned \cite{Al03} the use of the global store model is
identified as a potential source of \emph{incompleteness}.

\subsection{Technical contributions}

The proof system for object-oriented programs presented in our paper
is based on an assertion language comparable to JML. This allows for
the specification of dynamically evolving object structures at an
abstraction level which coincides with that of the programming
language: in this paper the only operations on objects we allow are
testing for equality and dereferencing.  Our transformation of the
considered object-oriented programs to recursive programs preserves
this abstraction level.  As a consequence we have to adapt existing
completeness proofs to recursive programs that use variables ranging
over \emph{abstract data types}, e.g., the type of objects.


In this paper we focus on \emph{strong} partial correctness which
requires absence of failures.
Note that absence of failures is naturally expressed by a corresponding condition on  the \emph{initial} state,
that is, by a  corresponding notion of \emph{weakest precondition}. Similarly, total
correctness of recursive programs is also naturally expressed by weakest
preconditions, see \cite{AB90}.  

To express weakest preconditions over abstract data types in an
assertion language \cite{TZ88} use a coding technique that requires a
weak second-order language. In contrast, we introduce here a new
state-based coding technique that allows us to express weakest
preconditions over abstract data types in the presence of infinite arrays
in a \emph{first-order} assertion language. 

Further, we generalize the original completeness proof of
\cite{Gore75} for the partial correctness of recursive programs to
weakest preconditions in order to deal with strong partial
correctness.  The completeness proof of \cite{Gore75} is based on the
expression of the \emph{graph} of a procedure call in terms of its
strongest postcondition of a precondition which ``freezes`` the
initial state by some fresh variables.  If we use instead weakest
preconditions to express the graph of a procedure call these freeze
variables are used to denote the \emph{final} state.  Because of
possible \emph{divergence} or \emph{failures} however we cannot eliminate in the
precondition these freeze variables by existential quantification.  As
such the completeness proof of \cite{Gore75} breaks down.  We show in
this paper how to restore completeness by the introduction of weakest
preconditions which explicitly model divergence and failures.

\subsection{Plan of the paper}

In the next section we introduce a kernel language that consists of
\textbf{while} programs augmented with the parallel assignment,
the failure statement and the block statement, and describe its
operational semantics.  In Section~\ref{sec:oo} we extend this kernel
language to a small object-oriented language that forms the subject of
our considerations.  In Section~\ref{OO:sec:sem} we define 
an operational semantics of this language.

Then, in Section \ref{sec:trans} we introduce a class of recursive
programs, define a transformation of the object-oriented programs to
recursive programs, and prove correctness of this transformation in an
appropriate sense.

Next, in Section \ref{sec:ass} we introduce the assertion language 
for object-oriented programs and
extend the substitution operation to instance variables.  In
Section \ref{sec:proof} we introduce the proof system that allows us
to prove correctness of the considered object-oriented programs.
Subsequently, in Section \ref{sec:formal} we explain how soundness and
relative completeness of this system can be established by reducing it
to the analysis of a corresponding proof system for recursive
procedures.

In Section \ref{sec:complete} we prove relative completeness of our
proof system for object-oriented programs on the basis of the
transformation, addressing the issues described above.  Finally, in
Section \ref{sec:extensions} explain how this approach can be extended
to deal with other features of object-oriented programming, like
classes, inheritance and subtyping, and with total correctness.  In
Appendix \ref{appendix-B} we list the rules defining the semantics of the kernel
language, the introduced proof rules and the introduced proof systems.

\section{Preliminaries}
\label{sec:prelim}

\subsection{A kernel language} 
\label{subsec:kernel}

We assume at least two {\it basic} types,
{\bf integer} and {\bf Boolean},
and for each $n \geq 1$ the 
{\it higher} types
$ T_1 \times \LL \times T_n \ra T$,
where  $T_1,\LL,T_n,T$ are basic types. 
$T_1,\LL, T_n$ are called {\it argument} types and 
$T$ the {\it value} type.
%
%
\emph{Simple} variables are of a basic type and \emph{array}
variables of a higher type.  By $V\!ar$ we denote the set of variable
declarations.  
Usually, we omit the typing information and
identify a variable declaration with the variable name.
%
Out of typed variables and typed constants
typed \emph{expressions} are constructed.
To deal with aliasing we use \emph{conditional} expressions
of the form $\ITE{B}{t_1}{t_2}$.
A \emph{subscripted variable} 
is an expression $a[t_1, \LL, t_n]$ 
for a suitably typed array variable $a$.

In this section we introduce the following small kernel programming language:
\[
\begin{array}{ll}
 S::= & skip \mid u:=t \mid 
         \bar{x}:=\bar{t} \mid 
         S_1;\ S_2 \mid 
         \ITE{B}{S_1}{S_2} \mid 
         \IF B \ra S_1 \ \FI \mid \\[1mm]
      &   \WDD{B}{S_1}
          \mid
        \block{\local \bar{x}:=\bar{t}; S_1}
\end{array}
\]
where $S$ stands for a typical statement or program, $u$ for a simple
or subscripted variable, $t$ for an expression (of the same type as
$u$), and $B$ for a Boolean expression.  Further, $\bar{x}:=\bar{t}$
is a parallel assignment, with $\bar{x} = x_1,\dots,x_n$ a non-empty
sequence of distinct simple variables and $\bar{t}=t_1,\dots,t_n$ a sequence
of expressions of the corresponding types.  The parallel assignment
plays a crucial role in our modeling of the parameter passing. 
%
The \textit{failure statement}
$\IF B \ra S_1 \ \FI$ is used to check the condition $B$ during the execution.
It raises a failure if $B$ is not satisfied. Thus it differs
from the abbreviation $\IT{B}{S} \equiv \ITE{B}{S}{skip}$.
%
%
%
To distinguish between local and global variables, we use
a \emph{block statement}
$\block{\local \bar{x}:=\bar{t}; S_1}$, 
where $\bar{x}$ is a 
non-empty sequence of simple distinct \emph{local}
variables, all of which are explicitly initialized by means of a
parallel assignment $\bar{x}:=\bar{t}$. 
We assume that the sets of local and global variables are disjoint.

%


For an expression $t$, we denote by $var(t)$ the set of all simple
and array variables 
in $t$.  Analogously, for a program
$S$, we denote by $var(S)$ the set of all simple and array variables 
in $S$, and by $change(S)$ the set of all global simple and array
variables that can be modified by $S$, i.e., the set of variables that
appear on the left-hand side of an assignment in $S$
outside of a
subscript position of a subscripted variable.

\subsection{$\LL$ and its semantics}
\label{subsec:kernel-sem}

We define the operational semantics of the kernel language in a
standard way, using a structural operational semantics in the sense of
Plotkin \cite{Plo04}.  
A \emph{configuration} $C$ is a pair $< S,\ 
\sg >$ consisting of a statement $S$ that is to be executed and a
\emph{state} $\sg$, i.e., a mapping that assigns to each variable
(including local variables) of type $T$ a value drawn from the set
${\cal D}_T$ denoted by type $T$.

Given a state $\sg$ and an expression $t$, we define in a standard way
its semantics $\sg(t)$, the value assigned to it by $\sg$.  Further, given a
sequence of expressions $\bar{t}$ (in particular, a sequence of
variables $\bar{x}$), we denote by $\sg(\bar{t})$ the corresponding
sequence of values assigned to $\bar{t}$ by $\sg$.

We denote the set of states by $\Sigma$.
Unless stated otherwise, the letters $\sg, \tau$ range over~$\Sigma$.
We use a \emph{special} state \FAIL\ to represent an abnormal situation in a program execution, a \emph{failure} in an execution of a program.
We stipulate that  $\FAIL \not\in \Sigma$. 
Sometimes to avoid confusion we refer to
the elements of $\Sigma$ as \emph{proper} states.

We use the notion of 
a \emph{state update} of a proper state~$\sg$, written as $\sg[u:=d]$, where $u$ is a simple or subscripted
variable of type $T$ and $d$ is an element of type $T$.
If $u$ is a simple variable then
$\sg[u:=d]$
is the state that agrees with~$\sg$ except for $u$ where its value is $d$.
If $u$ is a subscripted variable, say $u \equiv a[t_1,\LL,t_n]$, then
$\sg[u:=d]$
is the state that agrees with~$\sg$ except for the variable $a$ where
the value $\sg(a)(\sg(t_1),\LL,\sg(t_n))$ is changed to $d$.
For the special state we define the update
by ${\bf fail}[u:=d]={\bf fail}$.
Further, the parallel update $\sg[u_1,\ldots,u_n:=d_1,\ldots,d_n]$ of distinct
simple variables  is the state that agrees with~$\sg$ except for $u_i$ where its value is $d_i$,
for $i\in\{1,\ldots,n\}$.

A \emph{transition} is a step $C \ra C'$ between
configurations.  To express termination we use the empty statement
$E$; a configuration $< E,\ \sg >$ denotes termination in the state
$\sg$.  Transitions are specified by \emph{transition axioms} and
\emph{rules}.  The only transition axioms that are somewhat
non-standard deal with the block statement
and the failure statement. We write here
$\sg \Mo B$ to denote that $B$ is true in the state $\sg$.

\begin{itemize}
\item 
$<\IF B \ra S\ \FI,\sg> \ra <S,\sg>$, where $\sg \Mo B$, 

\vspace{2mm}

\item $<\IF B \ra S\ \FI,\sg> \ra <E,\FAIL>$, where $\sg \Mo \neg B$,

\vspace{2mm}

\item 
$<\block{\local \bar{x}:=\bar{t}; S},\sigma>\ra <\bar{x}:=\bar{t}; S;\bar{x}:=\sigma(\bar{x}),\sigma>$.

\end{itemize}
%

The last axiom ensures that the local variables are initialized as
prescribed by the parallel assignment and that upon termination the
local variables are
restored to their initial values, held at the beginning of the block
statement. This way we implicitly model a \emph{stack discipline}
for, possibly nested, blocks.
The other transition axioms and rules are standard (see
\ref{appendix-A}).

The {\it partial correctness semantics} is a mapping
$\MS{S}: \Sigma \ra {\cal P}(\Sigma)$ defined by
\[ 
 \MS{S}(\sg)=\C{\tau \in \Sigma \mid <S,\sg> \tra <E,\tau>},
\]
where $\tra$ denotes the reflexive, transitive closure of $\ra$.
The {\it strong partial correctness semantics} is a mapping
$\MSP{S}: \Sigma \ra {\cal P}(\Sigma \U \C{\FAIL} )$ defined by
\[ 
 \MSP{S}(\sg)=\C{\tau \in \Sigma \U \C{\FAIL} \mid <S,\sg> \tra <E,\tau>}. 
\]
So for all $S$ and $\sg$ we have $\FAIL \not\in \MS{S}(\sg)$,
while for some $S$ and $\sg$ we can have 
$\FAIL \in \MSP{S}(\sg)$.
In the latter case we say that $S$ \emph{can fail} when started in $\sigma$.
We extend these semantic mappings to \emph{sets} of states, $X \subseteq \Sigma$, 
by collecting all results obtained for the individual states $\sigma \in X$.


\section{Object-oriented programs: syntax}
\label{sec:oo}

To define the syntax of the considered object-oriented programming language
we  introduce a new basic type $\Object$ which denotes an
infinite set of objects $\Objects$.

\subsection{Expressions}
\label{subsec:exp}

An expression of type $\Object$
denotes an object.  Simple variables of type $\Object$ and array
variables with value type $\Object$ are called {\em object} variables.
We distinguish the simple object variable $\this$ which in each
state denotes the currently executing object.

Besides the set $V\!ar$ of variable declarations defined in Section \ref{sec:prelim} we now introduce a new set
$IV\!ar$ of instance variable declarations (so $V\!ar \cap IV\!ar=\emptyset$).  
An \emph{instance variable} can be a simple variable 
or an array variable.  Thus we now have two kinds of
variable declarations: the up till now considered declarations of
\emph{normal variables} ($V\!ar$),
which are shared, and the new declarations of instance variables ($IV\!ar$), which are owned by
objects. As before we identify each variable declaration with the variable name.
Out of instance array variables we construct, 
as in the case of normal array variables,
\emph{subscripted instance variables}.

For simplicity we assume that each object owns the same set of
instance variables.  Each object has its own local state which assigns
values to the instance variables.  We stipulate that $\this$ is a
normal variable, that is, $\this\in V\!ar$.

The only operation of a higher type which involves the basic type
$\Object$ (as argument type or as value type) is the equality
$=_\Object$ (abbreviated by $=$).  
Finally, we use the constant
$\nulll$ of type $\Object$ to represent the \emph{void reference},
a special construct which does not have a local state.


Summarizing, the set of expressions defined in Section~\ref{sec:prelim} is extended by the introduction of the basic type
$\Object$, the constant $\nulll$ of type $\Object$,
and the set $IV\!ar$ of (simple and
array) instance variables. Object expressions, i.e., expressions of
type $\Object$, can only be compared for equality.
A variable is either a normal variable (in
$V\!ar$) or an instance variable (in $IV\!ar$).  Simple
variables (in $V\!ar\cup IV\!ar$) can now be of type $\Object$. Also the
argument and the value types of array variables (in $V\!ar\cup IV\!ar$)
can be of type $\Object$.  Finally, we have the distinguished normal object variable
$\this$.

\subsection{Programs}
\mylabel{OO:sec:syntax}

For object-oriented programs we extend the syntax of the kernel language introduced in
Section~\ref{sec:prelim}. Assignments to instance variables are introduced as follows:
\[
  S::= u := t,
\]
where $u \in IV\!ar$ is a simple or subscripted (instance) variable.
Method calls are described by the clause
\[
  S::= s.m(t_1,\ldots,t_n),
\]
where $n \geq 0$.
Here the object expression $s$  denotes
the {\em called object}, the identifier $m$ denotes a method 
and $t_1,\ldots,t_n$ are the \emph{actual parameters}, which are
expressions of a basic type.
A method is defined by means of a \emph{declaration}
\[
  m(u_1,\ldots,u_n)::S,
\]
where the formal parameters $u_1,\ldots,u_n\in V\!ar$ are of a basic
type and $S$ is a statement called the \emph{method body}.  Since the
statements now include method calls, we allow for mutually
\emph{recursive} methods. 
However, the declarations cannot be nested, so we do not allow for
nested methods.

The instance variables appearing in the body $S$ of a method
declaration are owned by the executing object, which is denoted by the
variable $\this$.  To ensure correct use of the variable $\this$
we
disallow assignments to the variable $\this$.  
However, when describing
the semantics of method calls, we do use `auxiliary' block statements
in which the variable $\this$ {\em is} used as a local variable, so in
particular, it is initialized (and hence modified).
Further, to ensure that
instance variables are permanent, we require that in each block
statement 
instance
variables are not used as local variables.  

Apart from denoting the callee of a method call,  object expressions
can appear in  Boolean
expressions.  Further, we allow for assignments to object variables.

An \emph{object-oriented program} consists of a
\emph{main statement} $S$ built according to the syntax of this section and a given set $D$ of method declarations such that each method
used has a unique declaration in $D$ and each method call refers to a
method declared in~$D$.  
We assume that method calls are \emph{well-typed}, 
i.e., the numbers of formal and actual parameters agree and 
for each parameter position the types of the
corresponding actual and formal parameters coincide.
As before, name clashes between local variables and
global variables are resolved by assuming that 
no local variable of $S$ or $D$
occurs freely (i.e., as a global variable) in $S$ or $D$.
If $D$ is clear from the context we refer to the main statement
as an object-oriented program.

\begin{Example-HB} \mylabel{exa:find-object}
Consider the object-oriented program 
\[
 S \equiv \this.find(z)
\]
in the context of the recursive method declaration
\[
find(u) :: \IT{u \not= \this}{next.find(u)}.
\]
We assume that the actual parameter $z$, the formal parameter $u$, and
the instance variable $next$ are of type $\Object$.  The idea is that
$S$ checks whether a list of objects linked via the pointer $next$
contains an object stored in the actual parameter $z$. The search
through the list starts at the object stored in the variable $\this$.
\end{Example-HB}

\section{Object-oriented programs: semantics}\label{OO:sec:sem}

In this section we define the semantics of the introduced
object-oriented programs.  We first define the semantics of
expressions.  It requires an extension of the definition of state.
Subsequently we introduce a revised definition of an update of a state
and provide transition axioms concerned with the newly introduced
programming constructs.

\subsection{Semantics of expressions}
\label{subsec:sem-exp}
The main difficulty in defining the semantics of expressions is of
course how to deal properly with the instance variables. Each instance
variable has a different version (`instance') in each object.
Conceptually, when defining the semantics of an instance variable $u$
we view it as a variable of the form $\this.u$, where $\this$
represents the current object. So, given a proper state $\sigma$ and a
simple instance variable $x$ we first determine the current object
$o$, which is $\sigma(\this)$. Then we determine the \emph{local
  state}\index{state!local} of this object, which is $\sigma(o)$, or
$\sigma(\sigma(\this))$, and subsequently apply this local state to
the considered instance variable $x$. This means that given a proper
state $\sigma$ the value assigned to the instance variable $x$ is
$\sigma(o)(x)$, or, written out in full, $\sigma(\sigma(\this))(x)$.
This two-step procedure is at the heart of the definition of semantics
of an expression given below.

Next, we introduce a value
$\nulll\in \Objects$.  So in each proper state each variable of type
$\Object$ equals some object of $\Objects$, which can be the $\nulll$ object.  A proper
state $\sigma$ now additionally assigns to each object $o\in \Objects$
its local state $\sigma(o)$.
In turn, a local state $\sigma(o)$ of an object $o$ assigns a value of
appropriate type to each instance variable.
Note that by definition a proper state also assigns to $\nulll$ a local state.
However, by Lemma \ref{lem:safetyOO} from Section \ref{subsec:sem-prog} below
this state is not reachable in any computation.


Note that the local state of the current object $\sigma(\this)$ is
given by $\sigma(\sigma(\this))$.
Further, note that in particular, if an instance variable $x$ is of type $\Object$, then 
for each object $o \in \Objects$,
$\sigma(o)(x)$ is either
$\nulll$ or  an object $o'\in \Objects$, whose
local state is $\sigma(o')$, i.e., $\sigma(\sigma(o)(x))$. This
application of $\sigma$ can of course be nested, to get local states
of the form 
$\sigma(\sigma(\sigma(o)(x))(x))$, etc.

\begin{figure}[htbp]
\begin{center}
\includegraphics[width=0.8\textwidth]{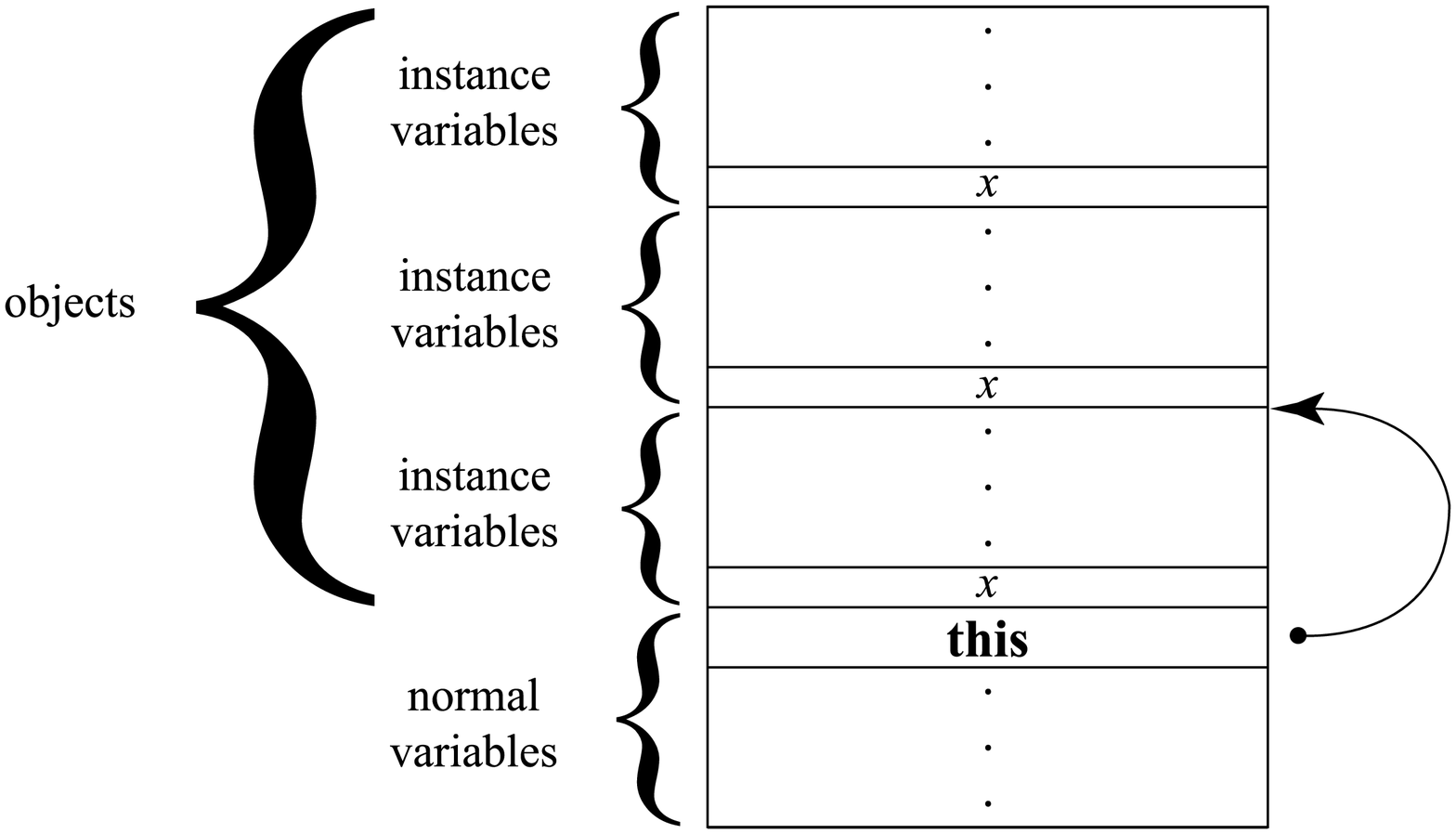}

\caption{\label{fig:heap} A state.}

\end{center}
\end{figure}

To illustrate the notion of a state consider Figure \ref{fig:heap}. The current object
is represented by a pointer to its memory region.  Each occurrence of
the variable $x$ is here an instance variable of a different object.
In contrast, the normal variables, in particular \textbf{this}, form
the global component of the state.

We need to extend the semantics $\sg(s)$ of an expression $s$ in a proper state $\sg$ 
(cf.~Section \ref{subsec:kernel-sem}) by the following clauses:

\begin{itemize}
\item if $s \equiv \nulll$ then $\sg(s)=\nulll$,
so the meaning of the constant $\nulll$ (representing the void reference) 
is the $\nulll$ object,

\item if $s \equiv x$ for some simple instance variable $x$ then
$\sg(s)=\sg(o)(x)$,
where $o=\sg(\this)$, so in expanded form this is
\begin{equation}
  \mylabel{equ:simple}
\sigma(x) = \sigma(\sigma(\this))(x). 
\end{equation}

\item if $s \equiv a[s_1,\ldots,s_n]$ for some instance array variable $a$ then

$$\sg(s)=\sg(o)(a)(\sg(s_1),\ldots,\sg(s_n)),$$
where $o=\sg(\this)$.
\end{itemize}

\subsection{Updates of states}
\label{subsec:update}
Next, we revise the definition of a state update
for the case of instance variables. 
Consider a proper state $\sigma$, a simple instance variable $x$,
and a value $d$ belonging to the type of $x$. To perform the
corresponding \emph{state update} of $\sigma$ on $x$ to $d$, written
as $\sigma[x:=d]$, we first identify the current
object $o$, which is $\sigma(\this)$ and its local state, which is
$\sigma(o)$, or $\sigma(\sigma(\this))$, that we denote by
$\tau$. Then we perform the appropriate update on the state $\tau$. 
So the desired update of $\sigma$ is achieved by modifying
$\tau$ to $\tau [x:=d]$.

In general, let $u$ be a (possibly subscripted) instance variable of type $T$ and
$\tau$ a local state. We define for $d\in{\cal D}_T$
$$
\tau[u:=d]
$$
analogously to the standard definition of state update
for normal variables. Furthermore,
we define for an object $o\in\Objects$ and local state $\tau$,
the state update $\sigma[o:=\tau]$ by
$$
\sigma[o:=\tau](o')=
\left\{
\begin{array}{ll}
\tau&\mbox{if}\ o=o'\\
\sigma(o')&\mbox{otherwise.}
\end{array}
\right.
$$

We  are now in a position to define the state update $\sigma[u:=d]$
for a (possibly subscripted) instance variable $u$ of type $T$ and
$d\in {\cal D}_T$, as follows:
$$
\sigma[u:=d]=\sigma[o:=\tau[u:=d]],
$$
where $o=\sigma(\this)$ and 
$\tau=\sigma(o)$.
Note that the state update $\sigma[o:=\tau[u:=d]]$ 
assigns to the current object $o$ the update
$\tau[u:=d]$ of its local
state $\tau$.
In its fully expanded form we get the following difficult to parse definition of a state update:
\[
\sigma[u:=d]=\sigma[\sigma(\this):=\sigma(\sigma(\this))[u:=d]].
\]

\begin{Example-HB}
Let $x$ be a Boolean instance variable,
$o=\sigma(\this)$, and $\tau=\sigma(o)$.
Then 
\begin{eqnarray*}
&   &\sigma[x:=\T](x)\\
& = & \quad \C{\mbox{(\ref{equ:simple}) with $\sigma$ replaced by   
        $\sigma[x:=\T]$}}\\
&   &\sigma[x:=\T](\sigma[x:=\T](\this))(x)\\
& = & \quad \C{\mbox{by the definition of state update, 
        $\sigma[x:=\T](\this)= \sigma(\this) = o$}}\\
&   &\sigma[x:=\T](o)(x)\\
& = & \quad \C{\mbox{definition of state update $\sigma[x:=\T]$}}\\
&   & \sigma[o:=\tau[x:=\T]](o)(x)\\
& = & \quad \C{\mbox{definition of state update 
        $\sigma[o:=\tau[x:=\T]]$}} \\
&   &\tau[x:=\T](x)\\
& = & \quad \C{\mbox{definition of state update $\tau[x:=\T]$}}\\
&   & \T.
\end{eqnarray*}
%
\end{Example-HB}

\subsection{Semantics of programs}
\label{subsec:sem-prog}

For the operational semantics of the considered programs we introduce
two transition axioms that deal with assignments to simple
or subscripted instance variables $u$ and with method calls
$s.m(\bar{t})$, where $\bar{t}$ is the list of actual parameters.  

\begin{itemize}
\item $<u:=t,\sigma>\ \rightarrow\ <E,\sigma[u:=\sigma(t)]>$, 
   
 \vspace{2mm}

\item $<s.m(\bar{t}),\sigma>\ \rightarrow\  <\IF s \not=\nulll \ra \block{\local \this,\bar{u}:=s,\bar{t};\ S} \ \FI,\sigma>$, \\[1mm]
where $m(\bar{u}):: S\in D$.

 

\end{itemize}


This clarifies that we use the stack discipline to handle the method
calls.  Indeed, the method body $S$ is executed in the state in which
the current object (denoted by the variable $\this$) becomes $\sg(s)$,
and upon termination of the method body $S$ the current object 
is restored to its previous value $\sg(\this)$ using the parallel assignment 
$\sg[\this,\bar{u}:=\sg(s,\bar{t})]$.
The use of the failure statement implies that
if in the considered state $\sigma$ the
called object $s$ equals the void reference (it equals $\nulll$), then
the method call yields a failure.

\begin{Lemma}\mytheoremname{Safety} \mylabel{lem:safetyOO}
For every statement $S$ that can arise during an execution of an object-oriented program and every proper state $\sg$, the following hold.

\begin{enumerate}[(i)]

\item \textbf{Absence of $\nulll$ Reference:}
if $\sigma(\this)\not=\nulll$ and $<S,\sg> \ra <S_1,\tau>$, then
$\tau(\this)\not=\nulll$.

\item \textbf{Type Safety:} 
   if $S$ is well-typed and $ <S,\sg> \ra <S_1,\tau>$
   holds, then also $S_1$ is well-typed.

\end{enumerate}

\end{Lemma}

\begin{Proof}
$(i)$ If $S \not\equiv E$ then any configuration $<S,\sg> $
has a successor in the transition relation $\ra$.
To prove the preservation of the assumed property of the state
it suffices to consider the execution of an  assignment $\this := s$.  
Each such assignment arises only within the context of the block statement in
the corresponding transition axiom  and is activated in a state $\sigma$
such that $\sigma(s) \neq \nulll$. This yields a state $\tau$ such that
$\tau(\this)\not=\nulll$.

\medskip

\noindent
$(ii)$
Except for method calls, the statements on the right-hand side of the
transition axioms are composed of the substatements of the statement on the left-hand
side of the transition axiom, which are well-typed by assumption.
Further, by the second transition axiom above, well-typed method calls lead to
well-typed parallel assignments in the block statements.
\end{Proof}

When considering verification of object-oriented programs we shall
only consider computations that start in a proper state $\sigma$ such
that $\sigma(\this) \neq \nulll$, i.e., in a state in which the current
object differs from the void reference.  The Safety Lemma
\ref{lem:safetyOO} implies that such computations never lead to a
proper state in which this inequality is violated.

The {\it partial correctness semantics} $\MS{S}$
and the {\it strong partial correctness semantics} $\MSP{S}$
of object-oriented programs $S$ are defined as for the kernel language.

\section{Transformation to recursive programs}
\label{sec:trans}

In this section we show that object-oriented programs
introduced in the previous section can be translated 
by means of a simple syntax-driven transformation to recursive programs with parameters.
Intuitively, for each method the current object is made into an explicit 
parameter of the corresponding recursive procedure.

\subsection{Recursive programs}
\label{subsec:rec}

As a preparation we introduce recursive programs
by adding recursive procedures with
call-by-value parameters to the kernel language. 
\emph{Procedure calls} with parameters are introduced by the grammar rule
\[
  S::= P(t_1,\ldots,t_n),
\]
where $P$ is a procedure identifier and $t_1,\ldots,t_n$, 
with $n \geq 0$, are expressions called \emph{actual parameters}. 
Procedures are defined by \emph{declarations} of the form
\[
  P(u_1,\ldots,u_n)::S,
\]
where $u_1,\ldots,u_n$ are distinct simple variables, called 
\emph{formal parameters} of the procedure $P$ and $S$ is the \emph{body} of the procedure $P$.

We assume a given set of procedure declarations $D$ such that each procedure that appears in $D$ has a unique declaration in $D$. 
A \emph{recursive program} consists of a \emph{main statement} $S$ built according to the syntax of this section and a given set $D$ of procedure declarations
such that all procedures whose calls
appear in the considered recursive programs are declared in $D$.
So we allow mutually recursive procedures but not nested procedures.
We assume that procedure calls are \emph{well-typed} in the same sense as method calls.
As in the case of the object-oriented programs,
name clashes between local variables and
global variables are resolved by assuming that no local variable of $S$
or $D$ occurs freely in $S$ or $D$.

\subsection*{Semantics}

For recursive programs we extend the operational semantics
of the kernel language by the following transition axiom 
that describes the \emph{call-by-value} parameter
mechanism.
\[
 <P(\bar{t}),\sigma> \ra 
          <\block{\local \bar{u}:=\bar{t};S},\sigma>, \ 
      \text{where}\ P(\bar{u}) ::S\in D.
\]
This yields for a recursive program $S$ the semantics $\MS{S}$ and $\MSP{S}$.

Note that thanks to the semantics of the block statement this axiom
correctly handles the clash between formal and actual parameters.
For example for $P(u) :: S \in D$ we get, as desired,
\[
 <P(u+1),\sg> \tra <S; u:=\sg(u), \sg[u:=\sg(u+1)]>.
\]

\subsection{Transformation}
\label{subsec:trans}

We now define a formal relation between object-oriented programs and
recursive programs.  We assume the class of recursive programs that
use normal variables whose type may involve the basic type $\Object$
and the class of object-oriented programs, as defined in Section
\ref{sec:oo}.  
Further, we assume for every declaration of an 
instance variable $u$ of a basic type
$T$ a declaration in {\it Var} of a normal array variable $u$ of type
$$\Object\ra T.$$
Similarly, we assume for every declaration of an 
instance variable $a$ of a higher type $T_1\times\ldots\times T_n\ra T$
a declaration in {\it Var} of a normal array variable $a$ of type
$$\Object\times T_1\times \ldots\times T_n \ra T.$$

A normal array variable of type
$$\Object\ra T$$
in the recursive program will represent 
the  instance variable of basic type
$T$ in the corresponding object-oriented program, and
a normal array variable of type
$$\Object\times T_1\times \ldots\times T_n \ra T$$
in the recursive
program will represent an instance variable of the corresponding
object-oriented program of type $T_1\times\ldots\times T_n\ra T$.

Given an `object-oriented' state $\sg$ we denote by
$\Theta(\sigma)$ the  `normal' state 
which represents the instance variables
as  normal variables. On normal variables 
of type $T$ the states $\sg$ and $\Theta(\sigma)$ agree and are of type
$V\!ar \ra {\cal D}_T$.
For instance variables 
of basic type $T$ the state $\sg$ is of type
\[
  \Objects \ra (IV\!ar \ra {\cal D}_T),
\]
the corresponding state $\Theta(\sigma)$ is of type
\[
 V\!ar \ra (\Objects \ra {\cal D}_T),
\]
and for instance array variables 
of type $T_1\times\ldots\times T_n\ra T$
the state $\sg$ is of type
\[
 \Objects \ra (IV\!ar \ra 
              ({\cal D}_{T_1} \times \ldots \times {\cal D}_{T_n} 
                                  \ra {\cal D}_T)),
\]
and the corresponding state $\Theta(\sigma)$ is of type
\[
 V\!ar \ra 
 (\Objects \times {\cal D}_{T_1} \times \ldots \times {\cal D}_{T_n} 
                                  \ra {\cal D}_T).
\]
Formally, $\Theta(\sigma)$ is defined as follows:
\begin{itemize}

\item 
$\Theta(\FAIL) = \FAIL$, 

\item
$\Theta(\sigma)(x) = \sigma(x)$, 
for every normal variable $x$,
\item
$\Theta(\sigma)(z)(o) = \sigma(o)(z)$,
for every object $o\in\Objects$ and normal array
variable $z$ of type $\Object \ra T$ on the left-hand side  of the equation
corresponding to an instance variable $z$ of a basic type $T$
on the right-hand side of the equation,
\item
$\Theta(\sigma)(a)(o,d_1,\ldots,d_n) = \sigma(o)(a)(d_1,\ldots,d_n)$,
for every object $o\in\Objects$ and  normal array variable $a$ of
type $\Object\times T_1\times\ldots\times T_n\ra T$ on the
left-hand side of the equation
corresponding to an instance array variable $a$ of 
type $T_1\times\ldots\times T_n\ra T$ on the right-hand side of the equation,
and $d_i\in {\cal D}_{T_i}$, for $i\in\{1,\ldots,n\}$.
\end{itemize}

Next, we define for every  expression
$s$ of the object-oriented programming language
the `normal' expression  $\Theta(s)$ of the recursive program
by induction on the structure of~$s$, with the following base cases:
\begin{itemize}
\item
$\Theta(x)\equiv x$, for every normal variable $x$,
\item
$\Theta(x)\equiv x[\this]$,
for every instance variable $x$ of a basic type,
\item
$\Theta(a[s_1,\ldots,s_n])\equiv
a[\this,\Theta(s_1),\ldots,\Theta(s_n)]$,
for every  instance array variable $a$.
\end{itemize}
The first case in particular yields $\Theta(\this)\equiv \this$.
The following lemma clarifies the outcome of this transformation.

\begin{Lemma}\mytheoremname{Translation} \mylabel{lem:transexpr}
For all proper states $\sigma$ the following hold.
\begin{enumerate}[(i)]

\item For all  expressions $s$, 
\[ 
 \sg(s) = \Theta(\sg)(\Theta(s)),
\]

where $\Theta(\sg)(\Theta(s))$ refers to the standard semantics of expressions
which involve only normal variables.

\item For all (possibly subscripted)
instance variables $u$ and values $d$ of the same type as $u$, 
\[ 
  \Theta(\sg[u:=d])=\Theta(\sg) [\Theta(u):=d].
\]
\end{enumerate}
\end{Lemma}

\begin{Proof}
By straightforward induction on the structure of
$s$ and case analysis on the structure of $u$.
\end{Proof}

Next, we extend by structural induction the transformation $\Theta$ 
to statements of the considered object-oriented language.
The failure statement is used to take care of the method calls 
on the void reference. We prove then that this transformation
preserves both partial and strong partial correctness semantics.

\begin{itemize}
\setlength{\itemsep}{1ex}

\item $\Theta(skip) \equiv skip$,

\item 
$\Theta(\bar{x}:=\bar{t})\equiv \bar{x} :=\Theta(\bar{t})$,

\item 
$\Theta(u:=s)\equiv \Theta(u):=\Theta(s)$,
\item
$\Theta(s.m(s_1,\ldots,s_n))\equiv 
\IF\ \Theta(s)\not=\nulll\ra\ m(\Theta(s),\Theta(s_1),\ldots,\Theta(s_n))\ \FI$,
\item
$\Theta(S_1;\ S_2)\equiv \Theta(S_1);\ \Theta(S_2)$,
\item
$\Theta(\ITE{B}{S_1}{S_2})\equiv \ITE{\Theta(B)}{\Theta(S_1)}{\Theta(S_2)}$,
\item
$\Theta(\WDD{B}{S})\equiv \WDD{\Theta(B)}{\Theta(S)}$,

\item
$\Theta(\IF B \ra S \ \FI)\equiv \IF \Theta(B) \ra \Theta(S) \ \FI$,

\item
$\Theta(\block{\local\ \bar{u}:=\bar{t}; S})\equiv
\block{\local\ \bar{u}:=\Theta(\bar{t}); \Theta(S)}$,

where $\Theta(\bar{t})$ denotes the result of applying $\Theta$
to the sequence of expressions $\bar{t}$.

\end{itemize}
So the translation of a method call $s.m(s_1,\ldots,s_n)$ transforms the called object
$s$ into an additional actual parameter of a call of
the procedure $m$. Additionally a check for a failure is added.
Finally, we transform every method declaration
$$
m(u_1,\ldots,u_n)::S
$$
into a procedure declaration
$$
m(\this,u_1,\ldots,u_n):: \Theta(S).
$$
So the distinguished normal variable $\this$
is added as an additional  formal parameter of the procedure $m$.
This way the set $D$ of method declarations is transformed
into the set 
$$
\Theta(D) = \{m(\this,u_1,\ldots,u_n):: \Theta(S)\mid\ 
m(u_1,\ldots,u_n):: S\in D\}
$$
of the corresponding procedure declarations.

\begin{Example-HB}
Consider the object-oriented program
\[
  S \equiv y.add(1);\ y.add(2),
\]
where $y$ is a normal variable of type \Object, 
in the context of the declaration
\[
D = \{add(x):: sum:=sum + x\},
\]
where the formal parameter $x$ is of type \textbf{integer}
and $sum$ is an instance variable, also
of type \textbf{integer}.
Then the transformation $\Theta$ yields
\[
 \Theta(S) \equiv 
  \IF\ y\not=\nulll\ra\ add(y,1)\ \FI;\
  \IF\ y\not=\nulll\ra\ add(y,2)\ \FI
\]
and
\[
 \Theta(D) = \{add(\this,x):: sum[\this]:=sum[\this] + x\}
\]
as the corresponding recursive program.
\end{Example-HB}

\subsection{Correctness proof}
\label{subsec:cor}

We have the following crucial correspondence between
an object-oriented program $S$ and its transformation
$\Theta(S)$.

\begin{Lemma}\mytheoremname{Transformation}\mylabel{oo:lem:corr}
For all well-typed object-oriented programs $S$, 
all sets of method declarations $D$,
all proper states $\sigma$,
and all proper or  $\FAIL$ states $\tau$,
$$
<S,\sg>\ \rightarrow^*\ <E,\tau>\ \mbox{\rm iff}\
<\Theta(S),\Theta(\sg)>\ \rightarrow^*\ <E,\Theta(\tau)>.
$$
\end{Lemma}

\begin{Proof}
We prove only the ($\Rightarrow$) direction. We
proceed by induction on the number
of the axiom and rule applications used in the computation
$<S,\sg>\ \rightarrow^*\ <E,\tau>$.

The only non-trivial case arises when $S$ begins with a method call, that is, is 
of the form $s.m(\bar{t}); S_1$. By the assumption, 
\[
<s.m(\bar{t}); S_1,\sg>\ \rightarrow \ <\block{\local \this,\bar{u}:=s,\bar{t};\ S}; S_1,\sigma> \
\rightarrow^*\ <E,\tau>,
\]
where $\sigma(s)\not=\nulll$ and $m(\bar{u}):: S\in D$.
So by the induction hypothesis and definition of $\Theta$,
\[
<\Theta(\block{\local \this,\bar{u}:=s,\bar{t};\ S}); \Theta(S_1),\Theta(\sigma)>
\rightarrow^*\ <E,\Theta(\tau)>.
\]
Note that
\[
\Theta(s.m(\bar{t}); S_1)  \equiv 
\IF\ \Theta(s)\not=\nulll\ra\ m(\Theta(s),\Theta(\bar{t}))\ \FI; \Theta(S_1),
\]
By the Translation Lemma \ref{lem:transexpr}$(i)$, 
we have $\Theta(\sigma)(\Theta(s))\not=\nulll$.
So by definition of the semantics of recursive programs and  definition of $\Theta$,
\[
\begin{array}{l}
\phantom{\ \rightarrow \ }<\IF\ \Theta(s)\not=\nulll\ra\ m(\Theta(s),\Theta(\bar{t}))\ \FI; \Theta(S_1), \Theta(\sigma)> \\[1mm]
\ \rightarrow^* \ 
<\Theta(\block{\local \this,\bar{u}:=s,\bar{t};\ S}); \Theta(S_1),\Theta(\sigma)>,
\end{array}
\]
which concludes the proof.
\end{Proof}

Finally, the following theorem establishes the correctness of the transformation $\Theta$ as a homomorphism.
We extend here $\Theta$ to a (possibly empty) set of states in an obvious way.

\begin{Theorem}\mytheoremname{Correctness of $\Theta$}
\mylabel{corr-transOO}
For all well-typed object-oriented programs $S$, 
all sets of method declarations $D$, and all proper states $\sigma$ 
the following holds:
\begin{enumerate}[(i)]
\item $\Theta(\MS{S}(\sigma))=\MS{\Theta(S)}(\Theta(\sigma))$,

\item $\Theta(\MSP{S}(\sigma))=\MSP{\Theta(S)}(\Theta(\sigma))$,
\end{enumerate}
where $S$ is considered in the context of the set $D$ 
and the corresponding recursive program $\Theta(S)$
in the context of the set of procedure declarations $\Theta(D)$.
\end{Theorem}

\begin{Proof}
The claim is a direct consequence of 
the Transformation Lem\-ma~\ref{oo:lem:corr}.
\end{Proof}

\section{Assertion language}
\label{sec:ass}

\subsection{Syntax and semantics} \label{subsec:oo-syntax}

Expressions of the programming language only refer to the local
state of the executing object and do not allow us to distinguish between different
versions of the instance variables. In the assertions we need to be more explicit.
So we introduce the set of {\em global expressions} which extends
the set of expressions of the object-oriented programming language
introduced in Section \ref{sec:oo} by the following additional clauses:
\begin{itemize}
\item
if $s$ is a global expression of type $\Object$
and $x$ is an instance variable of a basic type $T$ then
 $s.x$ is a global expression of type $T$,
\item
if $s$ is a global expression of type $\Object$,
$s_1,\ldots,s_n$ are global expressions of type $T_1,\ldots,T_n$, and $a$ is an
array instance variable of type $T_1\times\ldots\times T_n\rightarrow T$
then $s.a[s_1,\ldots,s_n]$ is a global expression of type $T$.
\end{itemize}
In particular, every expression of the programming language
is also a global expression.

\begin{Example-HB}
  Consider a normal integer variable $i$, a normal variable $x$ of type $\Object$,
a normal array variable $a$ of type ${\bf integer}\rightarrow\Object$, and 
an instance variable $next$ of type $\Object$.
Using them 
we can generate the following global expressions:
\[
  next,\ next.next,\ x.next,\ x.next.next,\ a[i].next,\ etc.,
\]
%
%
%
%
all of type $\Object$. In contrast, 
$next.x$ is not a global expression, since 
$x$ is not an instance variable.
\end{Example-HB}

We call a global expression of the form $s.u$  a \emph{navigation expression}
since it allows one to {\em navigate} through the local states of the  objects.
For example, the global expression $next.next$ refers to the object that can be reached by
`moving' to the object denoted by the value of $next$ of the current object $\this$
and evaluating the value of its variable $next$.

We define the semantics of global expressions 
by extending the semantics of expressions given in Section \ref{subsec:sem-exp}
as follows:
\begin{itemize}
\item
for a simple instance variable $x$ of type $T$,
$$\sg(s.x)=\sg(o)(x),$$
where $\sg(s) = o$,
%
\item
for an instance array variable $a$ with value type $T$,
$$
\sg(s.a[s_1,\ldots,s_n])= \sg(o)(a)(\sg(s_1),\LL,\sg(s_n)),
$$
where $\sg(s) = o$.
%
%
\end{itemize}


So for a simple or subscripted instance variable $u$ the semantics
of $u$ and $\this.u$ coincide, that is, for all proper states $\sigma$
we have $\sigma(u) = \sigma(\this.u)$.  In other words, we can view an
instance variable $u$ as an abbreviation for the global expression
$\this.u$.

Note that this semantics also provides meaning to global expressions
of the form $\nulll.u$. However, such expressions are meaningless when
specifying correctness of programs because the local state of the
$\nulll$ object can never be reached in computations starting in a
proper state $\sigma$ such that $\sigma(\this)\not=\nulll$ (see the
Safety Lemma~\ref{lem:safetyOO}).

\begin{Example-HB}
If $x$ is an object variable and $\sigma$ a proper state with
$\sigma(x)\not=\nulll$, then for all simple instance variables $y$
we have $\sigma(x.y)=\sigma(\sigma(x))(y)$.
\end{Example-HB}

\emph{Assertions} are constructed from global Boolean expressions
by adding quantification over simple \emph{normal}  variables. We use $p, q$
as typical letters for assertions.
For a state $\sg$ and an assertion $p$ we write
$\sg \Mo p$ if $\sg$ \emph{satisfies} $p$.
Let $\B{p}$ denote the set of proper states
satisfying~$p$, so $\B{p} = \{ \sg \in \Sigma \mid \sg \Mo p \}$.
So $\sg \Mo p$ iff $\sigma \in \B{p}$.

\subsection{Substitution and aliasing} \label{subsec:oo-subs}

We write $s[u:=t]$ for the result of 
substituting an expression $t$
for a simple or subscripted normal variable $u$ in an expression $s$. 
We call $[u:=t]$ a \emph{substitution}.
For a simple variable $u$ this is defined in the customary way.
Also it is straightforward how to define the simultaneous substitution
$s[\bar{x} := \bar{t}]$ involving a sequence of simple variables.

However, for a subscripted variable $u$, 
the problem of \emph{aliasing}, i.e., when syntactically different
subscripted variables denote the same location, has to be 
taken care of.
Following \cite{Bak80} we handle it using the conditional expressions.
For example, 
\[
min(a[x],y)[a[1]:=2] \equiv
      min(\ITE{x=1}{2}{a[x]}, y).
\]
The conditional expression checks whether $a[x]$ and $a[1]$
are \emph{aliases} of the same location. 
If so, the substitution of 2 for $a[1]$ results in $a[x]$ 
being replaced by 2,
otherwise the substitution has no effect.

Intuitively, in a given state $\sg$ the substituted expression
$s[u:=t]$ describes the same value as the expression $s$ evaluated in
the updated state $\sg[u:=\sg(t)]$, which arises after the assignment
$u:=t$ has been executed in $\sg$.  We shall later need the details of
the definition of $s[u:=t]$, so let us recall it here.  It proceeds by
induction on the structure of $s$.  The cases dealing with subscripted
variables are as follows:

\begin{itemize}

\item if $s \equiv a[s_1,\LL,s_n]$ for some array $a$, and
$u$ is a simple variable or a subscripted variable
$b[t_1,\LL,t_m]$ with $a \not\equiv b$, then
\[ s[u:=t] \equiv a[s_1[u:=t],\LL,s_n[u:=t]], \]

\item if $s \equiv a[s_1,\LL,s_n]$ for some array $a$ and
$u \equiv a[t_1,\LL,t_n]$ then
\[
  s[u:=t] \equiv \ITE{\bigwedge^{n}_{i=1}\ s'_i=t_i}
                     {t}{a[s'_1,\LL,s'_n]}
\]
where $s'_i\equiv s_i[u:=t]$ for $i \in \{1,\ldots,n\}$.

\end{itemize}

\NI
The most complicated case 
is the second clause for subscripted variables.
Here the conditional expression 
\[
 \ITE{\bigwedge^{n}_{i=1}\ s'_i=t_i}{\dots}{\dots}
\]
checks whether, for any given proper state $\sg$,
the expression $s \equiv a[s_1,\LL,s_n]$ in the updated state $\sg[u:=\sg(t)]$ and 
the expression $u \equiv a[t_1,\LL,t_n]$ in the state $\sg$ 
are aliases.
For this check the substitution $[u:=t]$ needs to applied 
inductively to all subscripts $s_1,\LL,s_n$ of $a[s_1,\LL,s_n]$. 
In case of an alias $s[u:=t]$ yields $t$. 
Otherwise, the substitution is applied inductively
to the subscripts $s_1,\LL,s_n$ of $a[s_1,\LL,s_n]$.

We now extend the definition of the outcome $s[u:=t]$ of the
substitution to the case of instance variables $u$ and global
expressions $s$ and $t$ constructed from them.  Let $u$ be a simple or subscripted
instance variable and $s$ and $t$ global expressions.  In general, the
substitution $[u:=t]$ replaces every possible alias $e.u$ of $u$ by
$t$.  In addition to the possible aliases of subscripted variables, we
now also have to consider the possibility that the global expression
$e[u:=t]$ denotes the current object $\this$. This explains the use of
conditional expressions below.

Here are the main cases of the definition of the substitution operation $s[u:=t]$:
\begin{itemize}
\item 
if $s\equiv x\in Var$ then
$$s[u:=t]\equiv s,$$
\item
if $s\equiv e.u$ and $u$ is a simple instance variable
then
$$
s[u:=t]\equiv \ITE{e'=\this}{t}{e'.u},
$$
where $e'\equiv e[u:=t]$,
\item
if $s\equiv e.a[s_1,\ldots,s_n]$ and $u\equiv a[t_1,\ldots,t_n]$ then
$$
s[u:=t]\equiv
 \ITE{e'=\this\wedge\bigwedge_{i=1}^n s'_i=t_i}{t}{e'.a[s'_1,\ldots,s'_n]},
$$
where $e'\equiv e[u:=t]$ and $s'_i\equiv s_i[u:=t]$ for $i \in \{1,\ldots,n\}$.
\end{itemize}

The following example should clarify this definition.

\begin{Example-HB}\mylabel{OO:ex:inst}
Suppose that $s \equiv \this.u$. Then
\begin{eqnarray*}
&        & \this.u[u:=t]\\
& \equiv & \ITE{\this[u:=t]=\this}{t}{\dots}\\
& \equiv & \ITE{\this=\this}{t}{\ldots}. 
\end{eqnarray*}
So $\this.u[u:=t]$ and $t$ are equal in the sense that  
for all proper states $\sigma$ we have
$\sg(\this.u[u:=t]) = \sg(t)$.

Next, suppose that $s \equiv \this.a[x]$, where $x$ is a simple variable. Then
\begin{eqnarray*}
&        & \this.a[x][a[x]:=t]\\
& \equiv & \ITE{\this[a[x]:=t]=\this \wedge x[a[x]:=t] = x}{t}{\dots}\\
& \equiv & \ITE{\this=\this \wedge x = x}{t}{\ldots}.
\end{eqnarray*}
So $\this.a[x][a[x]:=t]$ and $t$ are equal.
%
%
\end{Example-HB}

The substitution operation is then extended to assertions 
by properly taking care of quantification.
We have the following lemma that relates for instance variables
the effect of substitution to the state update.

\begin{Lemma}\mytheoremname{Substitution of Instance Variables}\mylabel{lem:assign-inst}
For all global expressions $s$ and $t$, all assertions $p$,
all simple or subscripted instance variables $u$ of the same type as $t$,
and all proper states $\sg$ the following hold:
\begin{enumerate}[(i)]
        \item $\sg(s[u:=t]) = \sg[u:=\sg(t)](s),$
        \item $\sg \Mo p[u:=t]$ iff $\sg[u:=\sg(t)] \Mo p$.
\end{enumerate}
\end{Lemma}

\begin{Proof}
By induction on the structure of $s$ and $p$.
\end{Proof}

\section{Proof theory for object-oriented programs}
\label{sec:proof}

We now study (strong) partial  correctness of object-oriented
programs expressed by
{\it correctness formulas} of the form $\HT{p}{S}{q}$,
where $S$ is a program and $p$ and $q$ are assertions. 
The assertion $p$ is the {\it precondition} of the correctness formula and $q$ is the {\it postcondition}.
A correctness formula \HT{p}{S}{q}\ holds in the sense of 
\emph{partial correctness}, abbreviated $\Mo \HT{p}{S}{q}$,
if every terminating computation of $S$ that starts in a state satisfying $p$ terminates in a state satisfying $q$.
And \HT{p}{S}{q}\ holds in the sense of 
\emph{strong partial correctness},
abbreviated $\Msp \HT{p}{S}{q}$, 
if $\Mo \HT{p}{S}{q}$
and no computation of $S$ that starts in a state satisfying $p$ ends in a failure.

Using the semantics $\cal M$ and ${\cal M}_{sp}$, we formalize 
these two interpretations of correctness formulas 
uniformly as set theoretic inclusions (cf.~\cite{ABO09}):

\begin{itemize}
\item $\Mo \HT{p}{S}{q}  \mbox{ if } \MS{S}(\B{p}) \sse \B{q}$,
\item $\Msp \HT{p}{S}{q}  \mbox{ if } \MSP{S}(\B{p}) \sse \B{q}$.
\end{itemize}
\NI
Since by definition $\FAIL \not\in \B{q}$ holds, 
$\MSP{S}(\B{p}) \sse \B{q}$
implies that $S$ does not fail
when started in a proper state $\sigma$ satisfying $p$,
as required for strong partial correctness.

\begin{Example-HB} 
Consider again the program $S \equiv \this.find(z)$
of Example \myref{exa:find-object}
for finding an object in a linked list.
To specify the desired effect of the there declared method $find$
we introduce a fresh normal array variable $a$ of type ${\bf integer}\ra \Object$ that  stores a linked list of objects, as expressed by
the assertion
\[
 p_0 \equiv \forall\, i \ge 0: a[i].next=a[i+1].
\]
We take 
$ p \equiv \this = a[0] \land p_0$
as precondition and 
$q \equiv \exists i\ge 0: z=a[i]$
as postcondition.
 Then the correctness
formula $\HT{p}{S}{q}$ holds in the sense of partial correctness,
i.e., upon termination $z$ will store one of the objects in the list.
Note that this is a correct specification since the variable $a$ is
not used (and hence not changed) in the program $S$.  In general,
normal auxiliary array or simple variables have to be used to record
the initial values of the program variables. 

However, this correctness formula does not hold in the sense
of strong partial correctness if the list contains the $\nulll$ object 
before the object stored in the variable $z$.
To avoid this we strengthen
the precondition by adding the assertion
\[
 p_1 \equiv\  \forall\, i \ge 0: a[i] \not=\nulll.
\]
%
%
Then $\HT{p\land p_1}{S}{q}$ holds 
in the sense of strong partial correctness.
Finally, if the list is circular and does not contain the $\nulll$
object or the object stored in $z$, the program $S$ diverges.
\end{Example-HB}

\subsection{Partial correctness} \label{subsec:rec-partial}

Partial correctness of the programs in the kernel language
is proved using the proof system {\it PK} consisting of the 
group of axioms and rules \myref{rul:skip}--\myref{rul:fail1},
and \myref{rul:block} shown in \ref{appendix-B1}.

We now consider partial correctness of object-oriented programs. 
First, we introduce the following axiom for assignments to 
instance variables:
\III

\NI
\textbf{AXIOM \myref{rul:ass-inst}:} ASSIGNMENT TO INSTANCE VARIABLES 
$$
\HT{p[u:=t]}{u:=t}{p}
$$
where $u$ is a simple or subscripted instance variable.
\III

So this axiom uses the new substitution operation defined in the
previous section.  Next, as we shall explain in a moment, we need the
following rule for weakening the precondition of a partial correctness
formula concerning a method call.  \III

\NI
\textbf{RULE \myref{rul:weak}:} WEAKENING 
\[
\frac{\HT{p\wedge s\not=\nulll}{s.m(\bar{t})}{q}}
{\HT{p}{s.m(\bar{t})}{q}}
\]

\subsection*{Non-recursive methods}

The main issue is how to deal with the parameters of method  calls.
Therefore, to focus on it we discuss the parameters of
non-recursive methods first.  The following {\em copy
  rule} shows how to prove correctness of
non-recursive method calls:

\[ 
 \frac{\HT{p}{\block{\local \this,\bar{u}:=s,\bar{t};\  S}}{q}}
        { \HT{p}{s.m(\bar{t})}{q}                          }
\]
where $m(\bar{u}) ::S\in D$\,.

\begin{Example-HB}
We prove the partial correctness formula
$
\HT{\T}{\nulll.m}{\F},
$
where $m::{\bf skip}\in D$. First, we have
$$
\HT{\F}{\block{\local \this:=\nulll;{\bf skip}}}{\F},
$$
so by the above copy rule we get
$
\HT{\F}{\nulll.m}{\F}.
$
The desired conclusion now follows 
by the above weakening rule and the consequence rule.
\end{Example-HB}

\subsection*{Recursive methods}

When we deal only with one recursive method and use the method
call as the considered object-oriented program, the above copy rule
needs to be modified to
\[ \frac{\HT{p}{s.m(\bar{t})}{q}  \vdash_{{\it PO}} \HT{p}{\block{\local \this,\bar{u}:=s,\bar{t};\  S}}{q}}
        { \HT{p}{s.m(\bar{t})}{q}                          }\]
where $D = \{m(\bar{u}) ::S\}$.
\III

The provability relation $\vdash_{{\it PO}}$ here refers to the 
proof system {\it PO}, which is defined as
\emph{PK} extended with the axiom~\myref{rul:ass-inst} for assignments to instance variables,
the weakening rule~\myref{rul:weak},
and the auxiliary rules~\ref{rul:disj}--\ref{rul:sub}
(as introduced in \ref{appendix-B2}).
Thus the premise of the rule states that in the proof the correctness
of the block statement we may \emph{assume} the corresponding
correctness formula concerning the method call.

In the case of an arbitrary program and a set of mutually recursive
method declarations we have the following generalization of the
above rule.  
\III

\NI
\textbf{RULE \myref{rul:rec1}:} RECURSION I
\[
\begin{array}{l}
\HT{p_1}{s_1.m_1(\bar{t}_1)}{q_1},\ldots,\HT{p_n}{s_n.m_n(\bar{t}_n)}{q_n} \vdash_{{\it PO}} \HT{p}{S}{q},                    \\
\HT{p_1}{s_1.m_1(\bar{t}_1)}{q_1},\ldots,\HT{p_n}{s_n.m_n(\bar{t}_n)}{q_n} \vdash_{{\it PO}} \\
\qquad \HT{p_i}{\block{\local \this,\bar{u}_i:=s_i,\bar{t}_i;\  S_i}}{q_i}, \ i \in \{1, \LL, n\} \\
[-\medskipamount]
\hrulefill                                                      \\
\HT{p}{S}{q} 
\end{array}
\]
where $m_i(\bar{u}_i) ::S_i\in D$ for $i \in \{1,\ldots,n\}$.
\III

The intuition behind this rule is as follows. Say that a program $S$
is $(p,q)$-\emph{correct} if $\HT{p}{S}{q}$ holds in the sense of
partial correctness.  The second premise of the rule states that we
can establish from the \emph{assumption} of the
$(p_i,q_i)$-correctness of the method calls $s_i.m_i(\bar{t}_i)$ for
$i \in \{1, \ldots, n\}$, the $(p_i,q_i)$-correctness of the procedure
bodies $S_i$ for $i \in \{1, \ldots, n\}$, which are adjusted as in
the transition axiom that deals with the method calls. Then we can
prove the $(p_i,q_i)$-correctness of the method calls
$s_i.m_i(\bar{t}_i)$ for $i \in \{1, \ldots, n\}$ unconditionally, and
thanks to the first premise establish the $(p,q)$-correctness of the
program $S$.

To prove partial correctness of object-oriented programs 
we use the following 

\medskip

\NI
PROOF SYSTEM ${\it PO}^+$ :

\medskip

\begin{minipage}{11cm}
This system is  obtained by extending 
{\it PO} by
the recursion I rule~\myref{rul:rec1}.
\end{minipage}
\III

\subsection{Strong partial correctness} 
%

Strong partial correctness of programs in the kernel language
is proved using the proof system {\it SPK} consisting of 
the group of axioms and rules \myref{rul:skip}--\myref{rul:cons}, 
\myref{rul:fail2}, and \myref{rul:block}
shown in \ref{appendix-B1}.

To prove strong partial correctness of method calls we modify the
above recursion rule I.  
The provability symbol $\vdash_{{\it SPO}}$ refers now
to the proof system {\it SPO}, which is defined as
\emph{SPK} augmented with the assignment axiom
\ref{rul:ass-inst}
and the
auxiliary rules \ref{rul:disj}--\ref{rul:sub} introduced in \ref{appendix-B2}.

%
%

\VV

\NI
\textbf{RULE \myref{rul:rec2}:} RECURSION II
\[
\begin{array}{l}
\HT{p_1}{s_1.m_1(\bar{t}_1)}{q_1},\ldots,\HT{p_n}{s_n.m_n(\bar{t}_n)}{q_n} \vdash_{{\it SPO}} \HT{p}{S}{q},                    \\
\HT{p_1}{s_1.m_1(\bar{t}_1)}{q_1},\ldots,\HT{p_n}{s_n.m_n(\bar{t}_n)}{q_n} \vdash_{{\it SPO}} \\
\qquad \HT{p_i}{\block{\local \this,\bar{u}_i:=s_i,\bar{t}_i;\  S_i}}{q_i}, \ i \in \{1, \LL, n\} \\
(*) \quad p_i \ra s_i\not=\nulll, \ i \in \{1, \LL, n\} \\
[-\medskipamount]
\hrulefill                                                      \\
\HT{p}{S}{q} 
\end{array}
\]
where $m_i(\bar{u}_i) ::S_i\in D$, for $i\in\{1,\ldots,n\}$.
\III

Thus compared with the recursion I rule \myref{rul:rec1},
the premises $(*)$ have been added. 
These premises are indeed needed,
as the following incorrect derivation shows.

\begin{Example-HB}
Let $m::{\bf skip}\in D$.
Without the premises $(*)$, we could derive from
\[
\HT{\T}{\nulll.m}{\T}\vdash \HT{\T}{\block{\local \this:=\nulll;{\bf skip}}}{\T}.
\]
the correctness formula
$\HT{\T}{\nulll.m}{\T}$.
%
However, this correctness formula
does not hold in the sense of strong partial correctness.
\end{Example-HB}

To prove strong partial correctness of object-oriented programs 
we use the following 

\medskip

\NI
PROOF SYSTEM ${\it SPO}^+$ :

\medskip

\begin{minipage}{11cm}
  This system is obtained by extending 
  {\it SPO} by
the recursion II rule~\myref{rul:rec2}.
\end{minipage}

%

\section{Formal justification}
\label{sec:formal}

To prove soundness and completeness of the proof systems \emph{PO} and
\emph{SPO} for (strong) partial correctness of object-oriented
programs we shall use the transformation given in Section
\ref{sec:trans}, notably the Correctness Theorem \ref{corr-transOO},
and reduce the problem to the analysis of the corresponding proof
systems for recursive programs.

The {\it partial correctness semantics} $\MS{S}$
and the {\it strong partial correctness semantics} $\MSP{S}$
of recursive programs $S$ are defined as for the kernel language.
We have
the following basic semantic invariance property of recursive programs.

\begin{lemma}\mytheoremname{Semantic Invariance} \mylabel{lem:seminv}
Let $\cal N$ stand for
$\cal M$ or ${\cal M}_{sp}$.
Further, let $\bar{z}$ be a sequence of fresh variables 
which  do not appear in the main statement
$S$ (or the given set of declarations $D$) and $\bar{d}$ be  a corresponding sequence of values.
Then
$$
\NS{S}(\sg[\bar{z}:=\bar{d}])=\{\tau[\bar{z}:=\bar{d}]\mid\ \tau \in \NS{S}(\sg)\}.
$$
\end{lemma}

\begin{Proof}
The proof proceeds by 
induction on the length of the computation.
\end{Proof}

\subsection{Proof theory for recursive programs}

Correctness formulas $\HT{p}{S}{q}$ for recursive programs
$S$ and their interpretation in terms of partial and strong partial correctness is defined as for object-oriented programs.

%
%
%

In the following rule for recursive programs
we use the provability symbol $\vdash$ to
refer to either the proof system \emph{PR} which consists of the proof system \emph{PK} augmented with the
auxiliary rules \ref{rul:disj}--\ref{rul:sub}
defined in \ref{appendix-B2} or the proof system \emph{SPR} which consists
of the proof system \emph{SPK} augmented with the these rules.
\III

\NI
\textbf{RULE \myref{rul:rec-proc}:} RECURSION III
\[
\begin{array}{l}
\HT{p_1}{P_1(\bar{t}_1)}{q_1},\ldots,\HT{p_n}{P_n(\bar{t}_n)}{q_n} \vdash \HT{p}{S}{q},                    \\
\HT{p_1}{P_1(\bar{t}_1)}{q_1},\ldots,\HT{p_n}{P_n(\bar{t}_n)}{q_n} \vdash \\
\qquad \HT{p_i}{\block{\local \bar{u}_i:=\bar{t}_i; S_i}}{q_i}, \ i \in \{1, \LL, n\} \\
[-\medskipamount]
\hrulefill                                                      \\
\HT{p}{S}{q} 
\end{array}
\]
where $P_i(\bar{u}_i) ::S_i\in D$.
\III

The intuition behind this rule is analogous as in the case of the recursion I rule
introduced in Section \ref{sec:proof}.
%
For recursive programs we use the following proof systems.
\III

\NI
PROOF SYSTEM ${\it PR}^+$ for partial correctness of
recursive programs:

\medskip

\begin{minipage}{11cm}
This system is  obtained by extending 
{\it PR}  by
the recursion~III rule~\myref{rul:rec-proc}.
\end{minipage}
\III

\NI
PROOF SYSTEM ${\it SPR}^+$ for strong partial correctness of
recursive programs:

\medskip

\begin{minipage}{11cm}
This system is  obtained by extending 
{\it SPR} by
the recursion~III rule~\myref{rul:rec-proc}.
\end{minipage}
\III


%
%
%

\subsection{Translation of assertions and proofs}

For the reduction to (correctness proofs of) recursive programs we
also have to transform expressions of the assertion language. 
To this end, we extend the definition of $\Theta(s)$ given in Section \ref{subsec:trans}
to global expressions introduced in Section \ref{subsec:oo-syntax}
by adding the following two cases (where $x$ is an instance variable of basic type
and $a$ is an array instance variable):

\begin{itemize}
\item $\Theta(s.x)=x[\Theta(s)]$,

\item $\Theta(s.a[s_1,\ldots,s_n])=a[\Theta(s), \Theta(s_1), \ldots, \Theta(s_n)]$.

\end{itemize}

Then we extend the transformation $\Theta(s)$ to
a transformation $\Theta(p)$ of assertions
by a straightforward induction on the structure of $p$.
Correctness of this transformation of assertions
is stated in the following lemma.

\begin{Lemma}\mytheoremname{Assertion}
\mylabel{lem:trans-ass}
For all assertions $p$ and all proper states $\sigma$
\[
\mbox{$\sigma \models p$ iff  $\Theta(\sigma) \models \Theta(p)$.}
\]
\end{Lemma}

\begin{Proof}
The straightforward proof proceeds by induction on the structure of
$p$.
\end{Proof}


\begin{Cor}\mytheoremname{Translation I}\mylabel{cor:OPtoRP}
For all correctness formulas $\HT{p}{S}{q}$, where $S$ is an object-oriented program,
\[
\mbox{ $\Mo \HT{p}{S}{q}$ iff $\Mo \HT{\Theta(p)}{\Theta(S)}{\Theta(q)}$,}
\]
and
\[
\mbox{ $\Msp{\HT{p}{S}{q}}$ iff $\Msp{\HT{\Theta(p)}{\Theta(S)}{\Theta(q)}}$.}
\]
\end{Cor}

\begin{Proof}
It follows directly by the Assertion Lemma \ref{lem:trans-ass} and
the Correctness Theorem \ref{corr-transOO}.
\end{Proof}

We next show that a correctness proof of an object-oriented program
can be translated to a correctness proof of the corresponding
recursive program.  We first need the following lemma which states
equivalence between a correctness proof of a method call from
a given set of assumptions and a 
correctness proof of the corresponding procedure call from the
translated set of assumptions.  For a given set of assumptions $A$
about method calls, we define the set of assumptions $\Theta(A)$ about
the corresponding procedure calls by
\[
\Theta(A) = \{ \HT{\Theta(p)}{m(\Theta(s),\Theta(\bar{t}))}{\Theta(q)} \mid \HT{p}{s.m(\bar{t})}{q}\in A \}.
\]


\begin{Lemma}\mytheoremname{Translation of Adaptation Correctness Proofs}\mylabel{lem:transmethodcalls}
Let $A$ be a given set of assumptions
about method calls.
Then
$$
A \vdash_{{\it AR}} \HT{p}{s.m(\bar{t})}{q} \ \mbox{ iff }
\Theta(A) \vdash_{{\it AR}} \HT{\Theta(p)}{m(\Theta(s),\Theta(\bar{t}))}{\Theta(q)},
$$
where $\vdash_{{\it AR}}$ denotes  provability in the proof system 
consisting of the so-called \emph{adaptation rules:}
the consequence rule \ref{rul:cons} and  the auxiliary 
proof rules introduced in \ref{appendix-B2}.

\end{Lemma}

\begin{Proof}
The proof proceeds by 
induction on the length of the derivation.
\end{Proof}

In order to prove the equivalence between
partial correctness proofs of a method call from a given set
of assumptions and correctness proofs of the corresponding procedure
call from the translated set of assumptions, we need the following lemma
about  partial correctness proofs of failure statements.

\begin{Lemma}\mytheoremname{Normal Form  Partial Correctness  Failure Statements}\mylabel{lem:normalformfail}
Let $A$ be a given set of assumptions
about procedure calls.
If
$$
A \vdash_{{\it PR}} \HT{p}{\IF B\ra S\ \FI}{q}
$$
then
$$
A \vdash_{{\it PR}} \HT{p\wedge B}{S}{q}.
$$
\end{Lemma}

\begin{Proof}
The proof proceeds by induction on the length of the given derivation.
By the form of the proof rules we can restrict to  the consequence rule \ref{rul:cons}, the auxiliary 
proof rules introduced in \ref{appendix-B2}, and the failure rule \ref{rul:fail1}.
We consider the case of an application of the auxiliary rule \ref{rul:intro}.
Let
$$
A \vdash_{{\it PR}} \HT{p'}{\IF B\ra S\ \FI}{q}
$$
and $p$ denote $\exists x:p'$,
where $x \not\in Var(D) \cup Var(S) \cup \mathit{free}(q)$.
By the induction hypothesis and an application of the auxiliary rule \ref{rul:fail1}, we have
$$
A \vdash_{{\it PR}} \HT{\exists x: (p'\wedge B)}{S}{q}.
$$
Since $x$ does not occur in $B$, the precondition is logically equivalent to $(\exists x:p')\wedge B$, so the desired result follows by an application of the consequence rule.
\end{Proof}

Next, we introduce the following lemmas stating the equivalence between
(strong) partial correctness proofs of a method call from a given set
of assumptions and correctness proofs of the corresponding procedure
call from the translated set of assumptions.


\begin{Lemma}\mytheoremname{Translation of Partial Correctness Proofs}\mylabel{lem:transpartial}
Let $A$ be a given set of assumptions
about method calls.
Then
$$
 A \vdash_{{\it PO}} \HT{p}{s.m(\bar{t})}{q}
$$
iff
$$
\Theta(A) \vdash_{{\it PR}} \HT{\Theta(p)}{ \IF \Theta(s)\not=\nulll \ra m(\Theta(s),\Theta(\bar{t}))\ \FI}{\Theta(q)}.
$$
\end{Lemma}

\begin{Proof}
Note that by the form of the proof rules we can restrict the rules of {\it PO} to
the proof system {\it AR} extended with the weakening rule \ref{rul:weak} and
restrict the rules of {\it PR} to the proof system {\it AR} extended with the 
failure rule \ref{rul:fail1}. 

\NI
$(\Rightarrow)$
We prove the claim by induction on the length of the derivation.
For the base case assume that $\HT{p}{s.m(\bar{t})}{q} \in A$. By
definition of $\Theta(A)$,
$$\HT{\Theta(p)}{m(\Theta(s),\Theta(\bar{t}))}{\Theta(q)}\in \Theta(A),$$
so
$$\Theta(A)\vdash_{{\it PR}} \HT{\Theta(p)}{m(\Theta(s),\Theta(\bar{t}))}{\Theta(q)}.$$
By a trivial application of the consequence rule, we get
$$
\Theta(A)\vdash_{{\it PR}} \HT{\Theta(p)\wedge \Theta(s)\not=\nulll}{m(\Theta(s),\Theta(\bar{t}))}{\Theta(q)}.
$$
Now by the failure rule, we get the desired result.


For the induction step we treat the case when the last rule applied is the weakening rule.
Then it is applied to
$$
A \vdash_{{\it PO}} \HT{p \wedge s\not=\nulll}{s.m(\bar{t})}{q}.
$$
By the induction hypothesis, 
$$
\Theta(A)\vdash_{{\it PR}} \HT{\Theta(p)\wedge \Theta(s)\not=\nulll}{\IF \Theta(s) \neq \nulll \ra m(\Theta(s),\Theta(\bar{t})) \ \FI}{\Theta(q)}.
$$
By Lemma \ref{lem:normalformfail}, it follows that
$$
\Theta(A)\vdash_{{\it PR}} \HT{\Theta(p)\wedge \Theta(s)\not=\nulll \wedge \Theta(s)\not=\nulll}{m(\Theta(s),\Theta(\bar{t}))}{\Theta(q)}, 
$$
so by the consequence and failure rules we get the desired result,
the right-hand side of the statement of the lemma.
\III

\NI
$(\Leftarrow)$
We prove the claim by induction on the length of the derivation.
%
We only treat the main case of the induction step when the last rule applied is the failure rule. 
Then it is applied to
$$
\Theta(A) \vdash_{{\it PR}}  \HT{\Theta(p)\wedge \Theta(s)\not=\nulll}{m(\Theta(s),\Theta(\bar{t}))}{\Theta(q)}.
$$
In this derivation in $\vdash_{{\it PR}}$ the failure rule has not been applied.
Thus we can replace $\vdash_{{\it PR}}$ by $\vdash_{{\it AR}}$.
By the Translation Lemma~\ref{lem:transmethodcalls}, we get
$$
A \vdash_{{\it AR}} \HT{p\wedge s\not=\nulll}{s.m(\bar{t})}{q}.
$$
Applying the weakening rule we get the desired result,
the left-hand side of the statement of the lemma.
\end{Proof}

\begin{Lemma}\mytheoremname{Translation of Strong Partial Correctness Proofs}\mylabel{lem:transtotal}
Let $A$ be a given set of assumptions
about method calls such that 
$p'\ra s'\not=\nulll$ holds for all
$\HT{p'}{s'.m'(\bar{t}')}{q'} \in A$.
Then
%
$$
A \vdash_{{\it SPO}} \HT{p}{s.m(\bar{t})}{q}
$$
iff
$$
\Theta(A) \vdash_{{\it SPR}} \HT{\Theta(p)}{\IF \Theta(s)\not=\nulll\ra m(\Theta(s),\Theta(\bar{t}))\ \FI}{\Theta(q)}.
$$

%
\end{Lemma}

\begin{Proof}

\NI
$(\Rightarrow)$
We prove the claim
by induction on the length of the derivation.
We only treat the base case, that is when $\HT{p}{s.m(\bar{t})}{q}\in A$.
By definition of $A$, the implication $p\rightarrow s\not=\nulll$ holds.
By definition of $\Theta(A)$,
$$\HT{\Theta(p)}{m(\Theta(s),\Theta(\bar{t}))}{\Theta(q)}\in \Theta(A).$$
Furthermore, by the Assertion Lemma \ref{lem:trans-ass}, we have
$\Theta(p)\rightarrow \Theta(s)\not=\nulll$.
So we conclude the desired result by an application of the failure II rule.
\III

\NI
$(\Leftarrow)$
We prove the claim by induction on the length of the derivation.
We only treat the main case of the inductive step, an application of the failure II rule. So 
$\Theta(p)\rightarrow \Theta(s)\not=\nulll$ holds and the rule is applied to
$$
\Theta(A) \vdash_{{\it SPR}}  \HT{\Theta(p)}{m(\Theta(s),\Theta(\bar{t}))}{\Theta(q)}.
$$
In this derivation in $\vdash_{{\it SPR}}$ the failure II rule has not been applied.
So we can replace $\vdash_{{\it SPR}}$ by $\vdash_{{\it AR}}$.
Thus by the Translation Lemma~\ref{lem:transmethodcalls}, 
$$
A \vdash_{{\it AR}} \HT{p}{s.m(\bar{t})}{q}.
$$
from which the desired result follows.
\end{Proof}

%
%
%
%
%

In order to extend the above lemmas from method calls to arbitrary statements
we need the following lemma which states that the transformation on assertions is a
homomorphism with respect to the substitution operation.

\begin{Lemma}\mytheoremname{Homomorphism}\mylabel{lem:homo}
For all global expressions or assertions $p$,  all expressions $t$ of the programming language, and all simple or subscripted variables $u$,
$$
\Theta(p[u:=t])\equiv \Theta(p)[\Theta(u):=\Theta(t)].
$$
\end{Lemma}

\begin{Proof}
We treat the case of a global expression $s$ and a simple instance variable $u$.
By definition,  $\Theta(u)\equiv u[\this]$.
It suffices to prove
$$
\Theta(s[u:=t])\equiv \Theta(s)[u[\this]:=\Theta(t)]
$$
by induction on the structure of the global expression $s$.
We treat the case of $s\equiv e.u$.
$$
\begin{array}{ll}
       &\Theta(e.u[u:=t]) \\[1mm]
\equiv & \quad \{\mbox{\rm by definition of the substitution $[u:=t]$}\}\\[1mm]
       &\Theta(\ITE{e[u:=t]=\this}{t}{e[u:=t].u}) \\[1mm]
\equiv & \quad\{\mbox{\rm by definition of $\Theta$}\} \\[1mm]
       & \ITE{\Theta(e[u:=t]=\this)}{\Theta(t)}{\Theta(e[u:=t].u)} \\[1mm]
\equiv & \quad\{\mbox{\rm by definition of $\Theta$}\} \\[1mm]
       & \ITE{\Theta(e[u:=t])=\this}{\Theta(t)}{u[\Theta(e[u:=t])]} \\[1mm]
\equiv & \quad\{\mbox{\rm by induction hypothesis about $e$}\}\\[1mm]
       &\ITE{\Theta(e)[u[\this]:=\Theta(t)]=\this}{\Theta(t)}{u[\Theta(e)[u[\this]:=\Theta(t)]]}\\[1mm]
\equiv & \quad\{\mbox{\rm by definition of the substitution $[u[\this]:=\Theta(t)]$}\}\\[1mm]
       & u[\Theta(e)][u[\this]:=\Theta(t)] \\[1mm]
\equiv & \quad\{\mbox{\rm by definition of $\Theta$}\} \\[1mm]
       & \Theta(e.u)[u[\this]:=\Theta(t)] 
\end{array}
$$
\end{Proof}

\begin{Lemma}\mytheoremname{Translation of Correctness Proofs Statements}\mylabel{lem:transstatement}
Let $A$ be a set of assumptions about method calls and 
$\HT{p}{S}{q}$ be a correctness formula of an object-oriented statement $S$.
Then
$$
A \vdash \HT{p}{S}{q}\;\mbox{ iff } \Theta(A)\vdash \HT{\Theta(p)}{\Theta(S)}{\Theta(q)},
$$
where 

\begin{itemize}

\item in case of partial correctness $\vdash$ on the left-hand side denotes  provability in the proof system
$PO$, and $\vdash$ on the right-hand side denotes 
provability in the proof system $PR$,
and 

\item in case of strong partial correctness
$\vdash$ on the left-hand side denotes  provability in the proof system
$SPO$, and $\vdash$ on the right-hand side denotes 
provability in the proof system 
$SPR$.
Additionally, we assume that $p'\ra s\not=\nulll$ holds for all
$\HT{p'}{s.m(\bar{t})}{q'} \in A$.

\end{itemize}

\end{Lemma}

\begin{Proof}
The proof proceeds by induction on the length of the derivation.
The case of an assignment statement follows by the Homomorphism Lemma \ref{lem:homo}.
The case of a method call follows 
by the Translation Lemma \ref{lem:transtotal}.
The cases of other program statements follow directly by
the induction hypothesis. In particular, 
in the cases of the consequence rule and the rules for conditionals
and loops, the Assertion Lemma \ref{lem:trans-ass} is used.
%
%
\end{Proof}

Finally, we arrive at the main result of this section.


\begin{Theorem}\mytheoremname{Translation II}\mylabel{cor:OPtoRP2}
For all correctness formulas $\HT{p}{S}{q}$, where $S$ is an object-oriented program,
\begin{enumerate}[(i)]
\setlength{\itemsep}{1ex}
\item
$\HT{p}{S}{q}$ is derivable in the proof system
{\it PO}$^+$ iff
$\HT{\Theta(p)}{\Theta(S)}{\Theta(q)}$
is derivable in {\it PR}$^+$, 
\item
$\HT{p}{S}{q}$ is derivable in the proof system
{\it SPO}$^+$ iff $\HT{\Theta(p)}{\Theta(S)}{\Theta(q)}$ is derivable in {\it SPR}$^+$.
\end{enumerate}
\end{Theorem}

\begin{Proof}
The proof proceeds by an induction on the length of the derivation.
The case of the assignment axioms is taken care of by the above Lemma \ref{lem:homo}.
The case of the recursion rules is taken care of by the Translation Lemma
\ref{lem:transstatement}.
The case of the other axioms and rules follows immediately from
the induction hypothesis (using the Assertion Lemma \ref{lem:trans-ass} in case of the rules for the conditional
and while statements).
Note that in the premises of the recursion rules we cannot apply the recursion rule again.
%
%
\end{Proof}


From the above theorem it immediately follows that the proof systems {\it PO}$^+$ and {\it SPO}$^+$ are
\emph{sound} and (relative) \emph{complete}  if and only if the corresponding proof systems {\it PR}$^+$ and {\it SPR}$^+$
are sound and (relative) complete.
For  proofs of soundness of the  systems {\it PR}$^+$ and {\it SPR}$^+$, that is,
for every correctness formula
$\HT{p}{S}{q}$ about a recursive program $S$,
derivability of $\HT{p}{S}{q}$ in  {\it PR}$^+$ and {\it SPR}$^+$ implies
$\Mo \HT{p}{S}{q}$ and $\Mo_{sp} \HT{p}{S}{q}$, respectively, we refer to our book \cite{ABO09}.
In the next section we discuss (relative) completeness of the proof systems {\it PR}$^+$ and {\it SPR}$^+$.
%
%
%


\section{Completeness}
\label{sec:complete}

We prove here relative completeness of the proof systems 
{\it PR}$^+$ and
{\it SPR}$^+$ for partial and strong partial correctness of the class of
recursive programs considered in this paper.  The proof is based on the
use of \emph{weakest preconditions}.  As explained in Section
\ref{sec:extensions}, this approach also applies to total correctness.  We first
discuss the expressibility of weakest preconditions for recursive
programs that use variables whose type may involve \emph{abstract data
  types} (like the basic type $\Object$).

\subsection{Expressibility}

We introduce the following definitions and conventions.  By $\sg=_V
\sg'$, for $V\subseteq {\it Var}$, we denote the fact that $\sg(v)=\sg'(v)$,
for $v\in V$.  We fix throughout this section a sequence
$\bar{x}=x_1,\ldots,x_k$ of (simple and array) variables and a main
statement $S$ such that its variables and those of the given set of
declarations $D$ are contained in $\bar{x}$.  Further, we fix a
corresponding sequence $\bar{y}$ of fresh variables used to refer to
the \emph{final} values of $\bar{x}$ in the definition of the weakest
preconditions below.  By $\bar{x}=\bar{y}$ we denote the conjunction
of the formulas $x_i=y_i$, for a simple variable $x_i$, and $\forall
\bar{u}_i: x_i[\bar{u}_i]=y_i[\bar{u}_i]$, for an array variable $x_i$
($\bar{u}_i$ denotes a sequence of simple variables corresponding to
the argument types of $x_i$).  The update
$\sg[\bar{x}:=\sg'(\bar{y})]$ assigns to each variable $x_i$ the
value/function $\sg'(y_i)$, for $i\in\{1,\ldots,k\}$. We denote by the
substitution $p[\bar{x}:=\bar{y}]$ the result of \emph{renaming} every
variable $x_i$ by $y_i$, for $i\in\{1,\ldots,k\}$.  We have the
following substitution lemma corresponding to the Substitution Lemma
\myref{lem:assign-inst}.

\begin{Lemma}\mytheoremname{Substitution} \mylabel{lem:renaming}
We have
$$
\mbox{$\sg\Mo p[\bar{x}:=\bar{y}]$ iff 
$\sg[\bar{x}:=\sg(\bar{y})]\Mo p$}.
$$
\end{Lemma}

\begin{Proof}
The proof proceeds by induction on the structure of~$p$.
\end{Proof}

The weakest precondition ${\it WP}(S,p)$ for  \emph{partial} correctness denotes the set
$$
\{\sg\mid\;  \MS{S}(\sg)\subseteq \B{p}\}.
$$
Similarly, the weakest precondition ${\it WP}_{sp}(S,p)$ for \emph{strong} partial correctness denotes the set
$$
\{\sg\mid\; \MSP{S}(\sg)\subseteq \B{p}\}.
$$

The above predicates satisfy the following equations.

\begin{Lemma}\mytheoremname{Weakest Precondition Calculus} \mylabel{lem:WP}
Let $W$ stand for ${\it WP}$ or ${\it WP}_{sp}$.
The weakest preconditions satisfy the following (standard) equations.
\begin{itemize}
\item
$W({\it skip},p)=\B{p}$,
\item
$W(u:=t,p)=\B{p[u:=t]}$,
\item
$W(\bar{x}:=\bar{t},p)=\B{p[\bar{x}:=\bar{t}]}$,
\item
${\it W}(S_1;S_2,p)={\it W}(S_1,{\it W}(S_2,p))$,
\item
${\it W}(\ITE{B}{S_1}{S_2},p)=(\B{B}\cap {\it W}(S_1,p))\cup (\B{\neg B}\cap {\it W}(S_2,p))$,
\item
${\it W}(\WDD{B}{S},p)=$

$(\B{\neg B}\cap\B{p})\cup (\B{B}\cap {\it W}( S,{\it W}(\WDD{B}{S},p)))$.
\end{itemize}
Failure statements  satisfy 
$${\it WP}(\IF B \ra S\ \FI,p)=(\B{B}\cap {\it WP}(S,p))\cup \B{\neg B}$$ 
and
$${\it WP}_{sp}(\IF B \ra S\ \FI,p)=\B{B}\cap {\it WP}_{sp}(S,p).$$
Finally, block statements satisfy 
$${\it W}(\block{\local \bar{u}:=\bar{t};S},p)=
{\it W}(\bar{u}:=\bar{t};S,p),
$$
where  the local variables $\bar{u}$ do not appear in $p$.
\end{Lemma}

\begin{Proof}
We prove the equation for block statements (the equations  for the other statements are standard).
By definition of the semantics of block statements and the above equations for (parallel) assignments and
sequential composition of statements, we have
$$
\begin{array}{ll}
   & {\it W}(\block{\local \bar{u}:=\bar{t};S},p)\\
=  & {\it W}(\bar{u}:=\bar{t};S;\bar{u}:=\sg(\bar{u}),p)\\
=  & {\it W}(\bar{u}:=\bar{t};S;{\it W}(\bar{u}:=\sg(\bar{u}),p))\\
=  & {\it W}(\bar{u}:=\bar{t};S,p).\\
\end{array}
$$
Note that $p[\bar{u}:=\sg(\bar{u})]$ equals $p$ because the variables $\bar{u}$ do not appear in $p$.
\end{Proof}

As a special case, we introduce the following \emph{most general weakest preconditions} 
$$
\mbox{${\it WP}(S,\bar{x}=\bar{y})$ and ${\it WP}_{sp}(S,\bar{x}=\bar{y})$}.
$$
Note that by definition,
$$
{\it WP}(S,\bar{x}=\bar{y})=\{\sg\mid\;  \MS{S}(\sg)\subseteq \{\sg[\bar{x}:=\sg(\bar{y})]\}\}
$$
and
$$
{\it WP}_{sp}(S,\bar{x}=\bar{y})=\{\sg\mid\; \MSP{S}(\sg)\subseteq \{\sg[\bar{x}:=\sg(\bar{y})]\}\}.
$$
These predicates  describe the \emph{graphs} of the \emph{deterministic} functions
$\MS{S}$ and $\MSP{S}$ in terms of a relation between the input variables $\bar{x}$ and
the output variables $\bar{y}$.

In order to express these most general weakest preconditions
in the \emph{first-order} assertion language we  introduce a \emph{state-based} encoding
of the basic types which allows for
a standard arithmetic encoding of the programming semantics.

Let {\bf nat} denote the basic type of the set $\NN$ of natural numbers.
For each basic type $T$  we fix a fresh array variable $h_T$ of type
${\bf nat}\ra T$ for a state-based encoding of the values of basic type
$T$. We will use $h$ to range over the variables $h_T$.
Without loss of generality we restrict our attention to states for which the interpretation
of each array variable $h_T$ specifies an \emph{enumeration} of the values of the basic type $T$,
that is, $h_T$ is \emph{surjective}, as expressed by
$$
\forall x:\exists n: x=h_T[n],
$$
where $x$ is of type $T$ and $n$ of type {\bf nat}.

Given this encoding of the basic types, we next show how to express in the assertion language
the encoding of the interpretation of the variables.
The assertion $code(n,z)$ defined by $h[n]=z$, where $z$ is a (simple) variable, directly  expresses that
the variable $n$ (of type {\bf nat}) stores an integer representation of  the value of $z$.
In order to express a similar assertion $code(n,a)$, where $a$ is an array variable, we assume in the assertion language
the following arithmetic operations:
\begin{itemize}
 \item
$<\bar{n}>$ denotes the natural number encoding the sequence of natural numbers $\bar{n}$,
\item
$n(i)$ denotes the $i$th element of the sequence encoded by $n$,
\item
$|n|$ denotes the length of the sequence of numbers encoded by $n$.
\end{itemize}
We note that the above operations can be formally defined in the assertion language by some \emph{computable} enumeration of all
finite sequences
of natural numbers (details are standard and therefore omitted).
Let $l$ denote the number of  argument types of
the array variable $a$.
The assertion $app(n,a)$ defined by
$$
|n|=l+1\wedge a[h[n(1)],\ldots,h[n(l)]]=h[n(l+1)]
$$
expresses that $n$ encodes an application of the interpretation of the array $a$.
The  assertion $code(n,a)$ is then defined by
$$ \bigwedge_{k=1}^{|n|}  app(n(k),a).$$
This assertion  expresses that $n$ encodes a finite sequence of numbers $n(k)$ each of which in turn
encodes an application of the interpretation of the array $a$.
For every  sequence $\bar{z}=z_1,\ldots,z_k$ of variables we denote by $code(n,\bar{z})$ the conjunction
$$
|n|=k\wedge \bigwedge_{i=1}^k code(n(i),z_i).
$$
Without loss of generality we assume that the  encoding of any sequence of  variables $\bar{z}$ is \emph{surjective}:
for every $n\in \NN$ there exists a state $\sg$ such $\sg\Mo code(n,\bar{z})$.

Given this  encoding of the interpretation of simple and array variables,
we next introduce the following binary \emph{arithmetic} relation ${\it comp}_S$, where $S$ is a recursive program,
which denotes the set
\begin{tabbing}
\qquad\qquad$\{(n,m)\mid$ \=$\forall \sg:\ $
\mbox{$\sg \models code(n,\bar{x})$ and $\sg\Mo code^+(m,\bar{x},\bar{y})$ implies} \\
\> $\sg[\bar{x}:=\sg(\bar{y})]\in\MS{S}(\sg)$\}.
\end{tabbing}
Here we implicitly assume that $code^+(m,\bar{x},\bar{y})$ asserts
$code(m,\bar{y})$ and \emph{additionally} enforces that
each array variable $x_i$  agrees with the corresponding array variable $y_i$
on the \emph{complement} of the domain specified by $m(i)$ (note that $m$ codes only a \emph{finite} part of each 
array variable of $\bar{y}$).
(The details of this extension of the assertion $code(m,\bar{y})$ are straightforward though somewhat tedious and therefore omitted.)
In the sequel we write $comp_S(n,m)$  to denote that  $n$ and $m$
belong to the binary relation $comp_S$ (we will use $n$ and $m$ both to denote natural numbers and variables of type {\bf nat}).


Since every finite  computation of $S$ accesses 
each array variable of $\bar{x}$ only on a finite subset of the domain of its interpretation,
we have the following closure property of ${\it comp}_S$.

\begin{lemma}\mytheoremname{Closure of {\it comp}} \mylabel{lem:clcomp}
For all states $\sg$ and $\sg'$ we have that $\sg'\in\MS{S}(\sg)$ implies
${\it comp}_S(n,m)$, for some pair of numbers $n$ and $m$ such that $\sg\Mo code(n,\bar{x})$ and $\sg[\bar{y}:=\sg'(\bar{x})]\Mo code^+(m,\bar{x},\bar{y})$.
\end{lemma}

We proceed with the introduction of
the (unary) arithmetic predicate ${\it fail}_S$ which denotes the set
$$
\{n \mid\ \forall\sg:\
\mbox{$\sg\Mo code(n,\bar{x})$ implies
 $\FAIL\in\MSP{S}(\sg)$}\}.
$$
In the sequel we also use $fail_S(n)$  to denote that $n$ is an element of the 
set $fail_S$.

Again, since every finite sequence of computation steps of $S$
accesses each array variable of $\bar{x}$ only on a finite subset of
the domain of its interpretation, we can assume the following closure
property of the ${\it fail}_S$.

\begin{lemma}\mytheoremname{Closure of {\it fail}} \mylabel{lem:clsp}
For every state $\sg$  such that $\FAIL\in\MSP{S}(\sg)$
we have
$\sg\Mo code(n,\bar{x})$ and ${\it fail}_S(n)$,  for some natural number $n$.
\end{lemma}

By means of standard  techniques for encoding  finite sequences of computation steps (see for example \cite{Bak80}) we
can express the predicates ${\it comp}_S$ and ${\it fail}_S$  arithmetically in the (first-order) assertion language
 in terms of the above  encoding of the interpretation of the variables
$\bar{x}$.
Therefore, we may assume without loss of generality that these predicates are present in the assertion language.

\begin{lemma}\mytheoremname{Expressibility} \mylabel{cor:exprwp}
For the assertion
$$
p\ \equiv\ \forall n, m: (code(n,\bar{x})\wedge {\it comp}_S(n,m))\ra code(m,\bar{y})
$$
we have
$$
{\it WP}(S,\bar{x}=\bar{y})=\B{p}
$$
and
$$
{\it WP}_{sp}(S,\bar{x}=\bar{y})=\B{p\wedge \forall n:code(n,\bar{x})\ra  \neg {\it fail}_S(n)}.
$$
\end{lemma}

\begin{Proof}
We prove first the first equation.
Let $\sg\in {\it WP}(S,\bar{x}=\bar{y})$, i.e.,
$\MS{S}(\sg)\subseteq \{\sg[\bar{x}:=\sg(\bar{y})]\}$.
In order to prove  $\sg\Mo p$, let
$\sg\Mo code(n,\bar{x})$ and  $comp_S(n,m)$, for some
arbitrary (constants) $n$ and $m$.
Further, let $\sg'\Mo code(m,\bar{y})$, for some $\sg'$ (note that the encoding of the
variables $\bar{y}$ is assumed to be surjective).
Without loss of generality we may assume that
$\sg[\bar{y}:=\sg'(\bar{y})]\Mo code^+(m,\bar{x},\bar{y})$
(note that $m$ only codes a finite part of $\sg'$).
Since the evaluation of $code(n,\bar{x})$ only depends on the interpretation of the variables $\bar{x}$,
$\sg\Mo code(n,\bar{x})$ implies  $\sg[\bar{y}:=\sg'(\bar{y})]\Mo code(n,\bar{x})$.
By definition of ${\it comp}_S$ it follows that
$\sg[\bar{x}:=\sg'(\bar{y})]\in\MS{S}(\sg)$ (note that $\bar{y}$ are assumed not to occur in $S$).
Because $S$ is deterministic it follows that $\sg[\bar{x}:=\sg'(\bar{y})]=\sg[\bar{x}:=\sg(\bar{y})]$,
i.e., $\sg'(\bar{y})=\sg(\bar{y})$.
So we conclude that $\sg\Mo code(m,\bar{y})$
(note that $code^+(m,\bar{x},\bar{y})$ trivially implies $code(m,\bar{y})$).

Next let $\sg\Mo p$.
In order to prove $\sg\in {\it WP}(S,\bar{x}=\bar{y})$, i.e.,
$\MS{S}(\sg)\subseteq \{\sg[\bar{x}:=\sg(\bar{y})]\}$, let $\sg'\in \MS{S}(\sg)$.
By Lemma \ref{lem:clcomp}, it follows that $comp_S(n,m)$, for some $n$ and $m$ such that
$\sg\Mo code(n,\bar{x})$ and $\sg[\bar{y}:=\sg'(\bar{x})]\Mo code^+(m,\bar{x},\bar{y})$.
So by the definition of ${\it comp}_S$ it follows that
$\sg[\bar{x}:=\sg'(\bar{x})]\in \MS{S}(\sg)$ (as above, note that $\sg\Mo code(n,\bar{x})$
implies $\sg[\bar{y}:=\sg'(\bar{x})]\Mo code(n,\bar{x})$ and the variables $\bar{y}$ do not appear in $S$).
Since $S$ is deterministic we conclude that $\sg'(\bar{x})=\sg(\bar{y})=\sg'(\bar{y})$.

For the second equation, it suffices to observe that by Lemma \ref{lem:clsp},
$\FAIL\in \MSP{S}(\sg)$ implies
$\sg\Mo \exists n: code(n,\bar{x})\wedge {\it fail}_S(n)$.
On the other hand, by definition of ${\it fail}_S$ it immediately follows that
$\sg\Mo \exists n:code(n,\bar{x})\wedge {\it fail}_S(n)$ implies
$\FAIL\in \MSP{S}(\sg)$.
We conclude that $\FAIL\not\in \MSP{S}(\sg)$ iff
$\sg\Mo  \forall n:code(n,\bar{x})\ra  \neg {\it fail}_S(n)$.
\end{Proof}

We conclude this discussion of the encoding of the most general weakest preconditions with the following
characterization of divergence
or failure, in case of partial correctness,
and divergence, in case of strong partial correctness.

\begin{Lemma}\mytheoremname{Expressibility of Divergence/Failure} \mylabel{lem:divfail}
For the assertion
$$
p\ \equiv\ \forall n, m: code(n,\bar{x})\ra \neg comp_S(n,m)
$$
we have
$$
{\it WP}(S,\F)=\B{p},$$
and
$$
{\it WP}_{sp}(S,\F)=\B{p\wedge \forall n:  code(n,\bar{x})\ra \neg fail_S(n)}.
$$

\end{Lemma}

The formula $p$ in the first equality expresses all states from which $S$ can diverge or
fail, while the formula on the right-hand side of the second equality expresses
all states from which $S$ can diverge.

\begin{Proof}
We prove the first equation (the second is dealt with as above).
First let $\sg\in {\it WP}(S,\F)$, i.e.,
$\MS{S}(\sg)=\emptyset$.
In order to prove  $\sg\Mo p$, let
$\sg\Mo code(n,\bar{x})$ and $comp_S(n,m)$, for some
arbitrary (constants) $n$ and $m$.
As above, we may assume without loss of generality that $\sg[\bar{y}:=\sg'(\bar{x})]\Mo code^+(m,\bar{x},\bar{y})$, for some $\sg'$.
By definition of ${\it comp}_S$, it follows  that
$\sg[\bar{x}:=\sg'(\bar{x})]\in\MS{S}(\sg)$
which contradicts $\MS{S}(\sg)=\emptyset$.

Next let $\sg\Mo p$.
In order to prove $\sg\in {\it WP}(S,\F)$, i.e.,
$\MS{S}(\sg)=\emptyset$,
let $\sg'\in \MS{S}(\sg)$.
By Lemma \ref{lem:clcomp}, it follows that $comp_S(n,m)$, for some $n$ and $m$ such that
$\sg\Mo code(n,\bar{x})$ (and  $\sg[\bar{y}:=\sg'(\bar{x})]\Mo code^+(m,\bar{x},\bar{y})$).
But this contradicts $\sg\Mo p$.
So we conclude that $\MS{S}(\sg)=\emptyset$.
\end{Proof}

\subsection{Completeness proof using most general correctness formulas}

We prove  (relative) completeness of the proof system  {\it SPR}$^+$, i.e, 
every strong partially correct specification $\HT{p}{S}{q}$
of a recursive program $S$ is derivable in {\it SPR}$^+$. Formally,
$\Mo_{sp}\HT{p}{S}{q}$ implies 
$\vdash_{{\it SPR}{^+}} \HT{p}{S}{q}$.
The proof of (relative) completeness of the proof
system {\it PR}$^+$ for partial correctness of recursive programs is similar.

First we state and prove the following completeness result for the most general correctness formulas.
Its formulation refers to the expressibility of the most general weakest preconditions justified by 
the Expressibility Lemma~\ref{cor:exprwp}.


\begin{Lemma}\mytheoremname{Completeness: Most General Correctness Formulas} \mylabel{lem:comp1}
Let $\bar{x}=x_1,\ldots,x_k$ be  all the variables (global and local) appearing in $D$, $S$, $p$ or $q$, and
$\bar{y}$ be  a corresponding sequence of fresh variables.
Further, let $q$ be a consistent assertion, i.e., $\B{q}\not=\B{\F}$.
We have 
$$
\mbox{$\Mo_{sp} \HT{p}{S}{q}$ implies $\HT{{\it WP}_{sp}(S,\bar{x}=\bar{y})}{S}{\bar{x}=\bar{y}}\vdash_{{\it SPR}} \HT{p}{S}{q}$.}
$$
\end{Lemma}

\begin{Proof}
Let $\Mo_{sp} \HT{p}{S}{q}$, with $q$ a consistent assertion.
Without loss of generality we may assume that $p$ and $q$ do not refer to the variables $\bar{y}$
(otherwise, we  rename them and apply the substitution rule).
Applying the invariance rule 
we obtain
$$
\HT{q[\bar{x}:=\bar{y}]\wedge {\it WP}_{sp}(S,\bar{x}=\bar{y})}{S}{q[\bar{x}:=\bar{y}]\wedge \bar{x}=\bar{y}}.
$$
Clearly the  postcondition implies $q$.
By the consequence rule, we then obtain
$$
\HT{q[\bar{x}:=\bar{y}]\wedge {\it WP}_{sp}(S,\bar{x}=\bar{y})}{S}{q}.
$$
Next we apply the  auxiliary rule \ref{rul:intro}:
$$
\HT{\exists\bar{y}:q[\bar{x}:=\bar{y}]\wedge {\it WP}_{sp}(S,\bar{x}=\bar{y})}{S}{q}.
$$
By definition of the weakest precondition and  $\Mo_{sp} \HT{p}{S}{q}$,
it follows that $p$ implies the above precondition 
(an application of the consequence rule thus gives us the desired 
correctness formula):
Let $\sg\Mo p$. 
It follows from $\Mo_{sp} \HT{p}{S}{q}$ that $\MSP{S}(\sg)\subseteq\B{q}$,
i.e., $\MSP{S}(\sg)=\emptyset$ or $\sg'\in \MSP{S}(\sg)$, for some proper state $\sg'$.
First we consider the case that $\MSP{S}(\sg)=\emptyset$.
From Lemma \ref{lem:seminv} it follows that
$\MSP{S}(\sg')=\emptyset$, for \emph{every} $\sg'$ such $\sg'=_{{\it Var}\setminus\bar{y}}\sg$.
Further, since $q$ is consistent, we have $\sg'\Mo q[\bar{x}:=\bar{y}]$,
for \emph{some} $\sg'$ such $\sg'=_{{\it Var}\setminus\bar{y}}\sg$.
Summarizing, we have
$$
\sg'\Mo q[\bar{x}:=\bar{y}]\wedge {\it WP}_{sp}( S,\bar{x}=\bar{y}),
$$
for some $\sg'$ such that $\sg'=_{{\it Var}\setminus\bar{y}}\sg$.
From which we conclude that
$$
\sg \Mo \exists\bar{y}:q[\bar{x}:=\bar{y}]\wedge {\it WP}_{sp}( S,\bar{x}=\bar{y}).
$$
Next we consider the case that $\sg'\in \MSP{S}(\sg)$, for some proper state $\sg'$.
From Lemma \ref{lem:seminv} it  follows that $\sg'\in \MSP{S}(\sg)$ implies
$\sg'[\bar{y}:=\sg'(\bar{x})]\in \MSP{S}(\sg[\bar{y}:=\sg'(\bar{x})])$.
Clearly, $\sg'[\bar{y}:=\sg'(\bar{x})]\Mo\bar{x}=\bar{y}$, and therefore
$\sg[\bar{y}:=\sg'(\bar{x})]\Mo {\it WP}_{sp}(S,\bar{x}=\bar{y})$.
Further, since $\bar{y}$ are assumed not to appear in $p$, $\sg\Mo p$ implies
$\sg[\bar{y}:=\sg'(\bar{x})]\Mo p$.
So we derive from the assumption $\Mo_{sp} \HT{p}{S}{q}$ that
$\sg'[\bar{y}:=\sg'(\bar{x})]\Mo q$.
By Lemma~\ref{lem:renaming} it then follows that  $\sg'[\bar{y}:=\sg'(\bar{x})]\Mo q[\bar{x}:=\bar{y}]$.
Since $\sg=_{{\it Var}\setminus \bar{x}}\sg'$ and the evaluation of $q[\bar{x}:=\bar{y}]$ does not depend
on the interpretation of the variables $\bar{x}$, we have
also $\sg[\bar{y}:=\sg'(\bar{x})]\Mo q[\bar{x}:=\bar{y}]$.
Summarizing, we obtain that
$$
\sg[\bar{y}:=\sg'(\bar{x})]\Mo q[\bar{x}:=\bar{y}]\wedge {\it WP}_{sp}(S,\bar{x}=\bar{y}),
$$
that is,
$$
\sg\Mo \exists\bar{y}:q[\bar{x}:=\bar{y}]\wedge {\it WP}_{sp}( S,\bar{x}=\bar{y}).
$$
\end{Proof}

We next introduce for  each  procedure call $P_i(\bar{t}_i)$, $i\in\{1,\ldots,n\}$, appearing in the given set of declarations $D$ or the main statement
$S$, the correctness formulas
$$
\mbox{ $\HT{{\it WP}_{sp}(P_i(\bar{t}_i),\F)}{P_i(\bar{t}_i)}{\F}$ and 
$\HT{{\it WP}_{sp}(P_i(\bar{t}_i),\bar{x}=\bar{y})}{P_i(\bar{t}_i)}{\bar{x}=\bar{y}}$},
$$
where $\bar{x}=x_1,\ldots,x_k$ are all variables (global and local) appearing in $D$ or $S$, and
$\bar{y}$ is a corresponding sequence of fresh variables.
We rely here on the expressibility of divergence and failure, as justified by 
the Expressibility Lemma~\ref{lem:divfail}.
Let $A$ denote the set of these correctness formulas.


\begin{lemma}\mytheoremname{Completeness Assumptions $A$} \mylabel{lem:comp2}
We have 
$$
\mbox{$\Mo_{sp} \HT{p}{S}{q}$ implies $A\vdash_{{\it SPR}} \HT{p}{S}{q}$.}
$$
\end{lemma}

\begin{Proof}
The proof proceeds by induction on the structure of the statement $S$.
Distinguishing between $\Mo_{sp} \HT{p}{S}{\F}$ and $\Mo_{sp} \HT{p}{S}{q}$, where $q$ is consistent,
by definition of ${\it WP}_{sp}(S,\F)$ and the above Lemma \ref{lem:comp1}, it suffices to prove
$$
A \vdash_{{\it SPR}} \HT{{\it WP}_{sp}(S,r)}{S}{r},
$$
where $r$ denotes the assertion $\F$ or $\bar{x}=\bar{y}$.
For assignments, sequential composition of statements, conditionals, failure statements and while 
statements the derivability of these correctness formulas follow from  the standard properties of
weakest preconditions as described in Lemma \ref{lem:WP}.
We consider therefore the non-standard case that $S$ denotes a block statement
$\block{\local \bar{u}:=\bar{t};S_1}$.
We introduce a sequence $\bar{z}$ of fresh variables corresponding
to the local variables $\bar{u}$.
By the semantics of block statements, it follows that
$$
\Mo_{sp}\HT{\bar{z}=\bar{u}\wedge {\it WP}_{sp}(S,r)}{ \bar{u}:=\bar{t};S_1}{r[\bar{u}:=\bar{z}]}.
$$
By the (general) induction hypothesis, we can derive this correctness formula from the
given set of assumptions $A$.
Next we  apply the block rule which gives
$$
\HT{\bar{z}=\bar{u}\wedge {\it WP}_{sp}(S,r)}{S}{r[\bar{u}:=\bar{z}]}.
$$
We proceed by an application of the invariance rule, which gives us
$$
\HT{\bar{z}=\bar{u}\wedge {\it WP}_{sp}(S,r)}{S}{\bar{z}=\bar{u}\wedge r[\bar{u}:=\bar{z}]}.
$$
The postcondition clearly implies $r$, so by the consequence rule, we obtain 
$$
\HT{\bar{z}=\bar{u}\wedge {\it WP}_{sp}(S,r)}{S}{r}.
$$
Finally, applying the substitution rule (replacing in the precondition $\bar{z}$ by
$\bar{u}$) followed a trivial application of  the consequence
rule gives us the desired result.
\end{Proof}

We conclude with the following main completeness theorem.

\begin{theorem}\mytheoremname{Completeness: Strong Partial Correctness} \mylabel{theo: completeness}
Every strong partially correct specification $\HT{p}{S}{q}$ of a 
recursive program $S$ is derivable in {\it SPR}$^+$.
Formally,
$\Mo_{sp}\HT{p}{S}{q}$ implies 
$ \vdash_{\mathit{SPR}{^+}} \HT{p}{S}{q}$. 
\end{theorem}

\begin{Proof}
Let $\Mo_{sp}\HT{p}{S}{q}$ and
$A$ be the set of assumptions as defined above.
By Lemma \ref{lem:comp2}, we have
$$
A \vdash_{\mathit{SPR}} \HT{p}{S}{q}.
$$
Next, let $r$ denote the assertion $\F$ or $\bar{x}=\bar{y}$.
We have that
$$
\Mo_{sp}\HT{{\it WP}_{sp}(P_i(\bar{t}_i)),r}{P_i(\bar{t}_i)}{r}
$$
implies
$$
\Mo_{sp}\HT{{\it WP}_{sp}(P_i(\bar{t}_i),r)}{\block{\local \bar{u}_i:=\bar{t}_i;S_i}}{r},
$$
for every $P_i(\bar{u}_i)::S_i\in D$.
By Lemma \ref{lem:comp2} again we have
$$
A\vdash_{\mathit{SPR}} 
\HT{{\it WP}_{sp}(P_i(\bar{t}_i),r)}{\block{\local \bar{u}_i:=\bar{t}_i;S_i}}{r},
$$
for every assumption of $A$.
Finally, by the recursion III rule \ref{rul:rec-proc}, we conclude that $\HT{p}{S}{q}$ is derivable in
the proof system {\it SPR}$^+$.
\end{Proof}

\section{Extensions}
\label{sec:extensions}

The approach to the verification of the object-oriented programs that
we proposed here is flexible and natural.  To substantiate this claim
we explain now how it can be naturally extended to other features of
object-oriented programming and to total correctness.

\subsection{Access to instance variables}

A natural possibility is to allow method calls to
access instance variables of arbitrary objects, a feature available in Java.
Then, given instance variables $x, y$ and object variables $s,t$,
we could use assignments such as $y := s.x + 1$, or $s.x := t.y + 1$,
and use global expressions in Boolean expressions, for example $2 \cdot s.x
= t.y + 1$, and as actual parameters in method calls, for example
$s.m(t.y + 1)$.

To extend the obtained results to the resulting programming language
the presentation would have to be modified in a number of places.
More precisely, such an extension requires the following:

\begin{itemize}
\item introduction of the global expressions already in Section~\ref{subsec:exp},

\item introduction of \emph{global terms}, which are expressions built out of global expressions
using the admitted function symbols (and respecting the well-typedness condition),

\item extension of the assignment statement to one of the form $s := t$,
where $s$ is a global expression and $t$ is a global term,

\item admission of the method calls of the form $s.m(t_1, \LL, t_n)$, where $s$ is an object expression
and $t_1, \LL, t_n$ are global terms,

\item introduction of the definition of semantics of global terms
in Section~\ref{subsec:sem-exp}, 

\item extension of the notion of an update of a state $\sg[s:=d]$ in 
Section~\ref{subsec:update}
to the case of a global expression $s$,

\item extension of the definition of substitution given in Section~\ref{subsec:oo-subs} to one of the form
$[s:=t]$, where $s$ is a global expression and $t$ is a global term,

\item extension of the transformation $\Theta$ given in Section~\ref{sec:trans}
to the considered programming language, by defining $\Theta(s)$ for a global expression $s$ already
in Section~\ref{subsec:trans},

\item extension of the assignment axiom~\ref{rul:ass-inst} to the
  above introduced class of assignments, and the recursion
  rules~\ref{rul:rec1} and~\ref{rul:rec2} to the above introduced
  method calls,

\item extension of the results of Section \ref{sec:formal}, notably 
the Homomorphism Lemma~\ref{lem:homo}, to this extended programming language.

\end{itemize}
The details are relatively straightforward and omitted.

\subsection{Object creation}

Most existing approaches to object creation (see for example in \cite{KeYBook2007}) 
follow implicitly the transformational approach by modeling it in terms of object \emph{activation}.
This can be made more explicit as follows.
Given an array variable {\it store} of type $\NN \ra \Object$ and a variable
{\it count} of type $\NN$, we model
the object creation statement $x:= \new$ by the statement
$$
{\it count}:= {\it count}+1; x:= {\it store}[{\it count}].
$$
This modeling of the object activation crucially depends on {\it store} being an \emph{unbounded} array variable.
Further it assumes that {\it store} is injective:
$$
\forall i:\forall j: i\not= j\ra {\it store}[i]\not= {\it store}[j],
$$
which is possible since we assumed that the type $\Object$ has infinitely many elements.

A drawback of this approach to object creation is that it involves
an explicit reference to a particular implementation.  
Since object variables can only be compared for equality or dereferenced, we show in Chapter~6 of our book \cite{ABO09}
 that we can in fact define a
\emph{substitution} $[x:=\new]$ which statically evaluates 
expressions in which $x$ occurs,
assuming that $x$ denotes a newly created object.  This in
turn allows us to define the weakest precondition $p[x:=\new]$ of the
object creation statement $x:=\new$ and w.r.t.~a postcondition $p$ which
abstracts from the particular implementation of object creation.
This yields the assignment statement
\[ \HT{p[x:=\new]}{x:=\new}{p} \]
allowing us to reason about object creation.

\subsection{Classes and Inheritance}
The transformational approach discussed in this paper can be readily
extended to deal with various  features of mainstream object-oriented
languages, like classes, inheritance and polymorphism (i.e.,
subtyping). 
As an example, 
we now discuss the details of such a
transformation for a fragment of Java that extends the object-oriented language
considered so far with {\em dynamic binding of methods}.
This extension comprises the following:

\begin{itemize}

\item introduction of \emph{classes} as basic types, 
  
\item use within the context of each program of a reflexive and
  transitive \emph{subclass relation} and its inverse \emph{superclass
    relation} defined on the set of classes used;
we assume that this relation respects \emph{single inheritance}, i.e.,
each class has at most one direct superclass, and that $\Object$ is
the superclass of each class,

 \item introduction of the assignment $u:=t$, where the type
   of the object expression $t$ is a subclass of the type of $u$, and of the method call
   $s.m(t_1,\ldots,t_n)$, where for $i\in\{1,\ldots,n\}$ the type of the actual
   parameter $t_i$ is a subclass of the type of the
      corresponding formal parameter,  in case  $t_i$ is an object expression,
\item introduction of mutually disjoint sets ${\cal D}_C\subseteq \Objects$ of object instances of class~$C$,

\end{itemize}

Further, we associate with each class a set of method declarations.
The instance variables of a class $C$ are the inherited ones plus the
ones that are introduced in the method declarations of $C$.
An object-oriented program in this new setting consists then of a main statement and a set of classes, each
with its set of method declarations, and a subclass relation.
The semantics of a method call in this extension is captured by the rule
$$<s.m(\bar{t}),\sigma>\ \rightarrow\  <\IF s \not=\nulll \ra \block{\local \this,\bar{u}:=s,\bar{t};\ S \ \FI},\sigma>,$$
where $S$ is such that $\sigma(s)\in {\cal D}_C$ and $m(\bar{u}):: S\in C$.
In words, the class of the object denoted by  the  expression $s$ determines the actual definition of the called method to be used.
Note that this class in general is a subclass of the type of the  expression $s$.
%

We now explain how the programs formed in this extended setting can be transformed to the programs
considered earlier extended by
an introduction for each class  $C$ of a unary predicate $C: \Object\ra {\bf Boolean}$ whose semantics is
defined by
$$
\sigma(C(s))=\left\{
\begin{array}{ll}
\T & \mbox{\rm if $\sigma(s)\in {\cal D}_C$}\\

\F& \mbox{\rm otherwise.}
\end{array}
\right.
$$

In order to model dynamic binding in this  extended object-oriented language
we first {\em flatten} the  inheritance hierarchy between classes  by introducing
a global set of method definitions $D$ which consists of all method definitions 
$$m@C(\bar{u})::S,$$
 where the declaration $m(\bar{u})::S$ appears in the class $C$ itself or 
in the `minimal' superclass $C'$ of $C$, that is,
no other superclass of $C$ which is also a subclass of $C'$ contains a declaration of $m$.
%
%
%
We then model the semantics of a method call $s.m(\bar{t})$ by the  statement $S_n$,
inductively defined for $i\in\{0,\ldots,n-1\}$ by 
$$
\begin{array}{ll}
S_0 & \equiv {\it skip}\\
S_{i+1} & \equiv \IF C_{i+1}(s) \THEN s.m@C_{i+1}(\bar{t}) \ELSE S_i \ \FI,
\end{array}
$$
where $\{C_1,\ldots,C_n\}$ is the set of subclasses of  the type of $s$.

After establishing an analogue of Theorem \ref{corr-transOO} for
the above transformation one could verify the programs written
in the source language by verifying their translated version. In
principle one could also derive proof rules that deal with the source
programs directly, analogously as in Section \ref{sec:proof}.

\subsection{Total correctness}

To focus on the crucial aspects of our approach to verification we did
not deal with program termination.  The appropriate extension combines
strong partial correctness with termination and requires the following:

\begin{itemize}
\item addition of a special state $\bot$ that models divergence,

\item modification of the definition of semantics to take care of divergence,

\item introduction of a new notion of soundness of a proof system,

\item replacement of the current LOOP rule \ref{rul:loop}
by a rule that also takes care of termination,

\item replacement of the current the recursion rules~\ref{rul:rec1} and ~\ref{rul:rec2}
by a single rule that also takes care of termination, 

\item similarly for the recursion~III rule~\ref{rul:rec-proc},

\item appropriate modification of the proofs in 
Section \ref{sec:formal} to additionally deal with termination.

\end{itemize}
The details are presented in \cite[chapter 6]{ABO09}.
Since termination is, roughly speaking, orthogonal to object-orientation, 
the transformational approach for (strong)
partial correctness can be extended to total correctness in a straightforward, though somewhat tedious, manner.

\section{Conclusion}
\label{sec:conc}

We presented here an assertional proof system to reason about partial
and strong partial correctness of a class of object-oriented programs.
Its formal justification (that is, soundness and relative
completeness) was carried out using a syntax-directed transformation to
recursive programs.  

We proved a new relative completeness result for a class of recursive
programs that use variables ranging over abstract data types (like the
basic type \Object) and showed that the transformation preserves
completeness.  We also showed that the transformational approach can
be applied to intricate and complex object-oriented features, such as
inheritance and subtype polymorphism, by transforming them in the
context of a {\em closed} program to the core language considered in
this paper.

%

Extension of the transformational approach to
{\em open} object-oriented programs, so programs that do not necessarily
include the definitions of all the classes used (in Java for example such classes
are imported from {\em packages}),
however, requires an additional study of
structuring recursive programs by means of {\em modules} along
the lines of the Modula programming language \cite{Wirth89} and of the
corresponding proof-theoretical concept of a {\em contract} as
introduced in the Eiffel programming language \cite{Eiffel}.

\appendix

%

\section{Semantics}
\label{appendix-A}

In the following we list the omitted transition axioms and rules that define the transition relation~$\ra$.

\begin{enumerate}[(i)]
\setlength{\itemindent}{0mm}
\setlength{\itemsep}{1ex}

\item \mylabel{trn:1}
$<skip,\sg> \ra <E,\sg>$,

\item \mylabel{trn:2}
$<u:=t,\sg> \ra <E,\sg[u:=\sg(t)]>$, \\[1mm]
where $u\in Var$ is a simple variable
or $u\equiv a[s_1,\ldots,s_n]$, for $a\in Var$,

\item \mylabel{trn:2a}
$<\bar{x}:=\bar{t},\sg> \ra <E,\sg[\bar{x}:=\sg(\bar{t})]>$

\vspace{1mm}

\item \mylabel{trn:3}
$\displaystyle\frac{\raisebox{1mm}{$<S_1,\sg> \ra <S_2,\tau>$}}
             {\raisebox{-1mm}{$<S_1;\ S,\sg> \ra <S_2;\ S,\tau>$}}$

\vspace{1mm}

\item \mylabel{trn:4}
$<\ITE{B}{S_1}{S_2},\sg> \ra <S_1,\sg>$, where $\sg \Mo B$,

\item \mylabel{trn:5}
$<\ITE{B}{S_1}{S_2},\sg> \ra <S_2,\sg>$, where $\sg \Mo \neg B$,

\item \mylabel{trn:6}
$<\WDD{B}{S},\sg> \ra <S;\ \WDD{B}{S},\sg>$, where $\sg \Mo B$,

\item \mylabel{trn:7}
$<\WDD{B}{S},\sg> \ra <E,\sg>$ where $\sg \Mo \neg B$.

\end{enumerate}

\section{Axioms and proof rules}
\label{appendix-B}

In the following we list the used axioms and proof rules.
Given an assertion $q$ we denote below its set of free variables by $\mathit{free}(q)$.


\subsection{Axioms and proof rules for the kernel language}
\label{appendix-B1}

To establish correctness of programs from the kernel language of Section \ref{sec:prelim}
we rely on the following axioms and proof rules.

\begin{Axiom} SKIP \mylabel{rul:skip}
\[ \HT{p}{skip}{p} \]
\end{Axiom}

\begin{Axiom} ASSIGNMENT \mylabel{rul:assi}
\[ \HT{p[u:=t]}{u:=t}{p} \]
where $u\in V\!ar$ or $u\equiv a[s_1,\ldots,s_n]$ and $a\in V\!ar$.
\end{Axiom}

\begin{Axiom} PARALLEL ASSIGNMENT \mylabel{rul:assip}
\[ 
 \HT{p[\bar{x}:=\bar{t}]}{\bar{x}:=\bar{t}}{p} 
\]
\end{Axiom}

\begin{Rule} COMPOSITION \mylabel{rul:comp}
\[ \frac{ \HT{p}{S_1}{r}, \HT{r}{S_2}{q}                }
        { \HT{p}{S_1;\ S_2}{q}                          }\]
\end{Rule}

\begin{Rule} CONDITIONAL \mylabel{rul:cond}
\[ \frac{ \HT{p \A B}{S_1}{q}, \HT{p \A \neg B}{S_2}{q}         }
        { \HT{p}{\ITE{B}{S_1}{S_2}}{q}                          }\]
\end{Rule}

\begin{Rule} LOOP \mylabel{rul:loop}
\[ \frac{ \HT{p \A B}{S}{p}                             }
        { \HT{p}{\WDD{B}{S}}{p \A \neg B}               }\]
\end{Rule}

\begin{Rule} CONSEQUENCE \mylabel{rul:cons}
\[ \frac{ p \ra p_1, \HT{p_1}{S}{q_1}, q_1 \ra q        }
        { \HT{p}{S}{q}                                  }\]
\end{Rule}


\begin{Rule} \mylabel{rul:fail1} FAILURE
\[ \frac{ \HT{p \A B}{S}{q}        }
        { \HT{p}{\IF B \ra S\ \FI}{q}}
\]
\end{Rule}

\begin{Rule} \mylabel{rul:fail2} FAILURE II
\[ \frac{ p \ra B, \HT{p}{S}{q}        }
        { \HT{p}{\IF B \ra S\ \FI}{q}                          }\]
\end{Rule}

\begin{Rule} \mylabel{rul:block} BLOCK
\[
\frac{\HT{p}{\bar{x} := \bar{t};\  S}{q}}
{\HT{p}{\block{\local \bar{x} := \bar{t};\  S}}{q}}
\]
where $\C{\bar{x}} \cap \mathit{free}(q) = \ES$.
\end{Rule}

\subsection{Auxiliary rules}
\label{appendix-B2}

Further, we rely on the following auxiliary axioms and proof rules
that occasionally refer to the assumed set of procedure or method declarations $D$.
We refer in them to the sets of variables $var(D)$ and $change(D)$ defined in the expected way.


\begin{AuxRule} DISJUNCTION \mylabel{rul:disj}
\[ \frac{ \HT{p}{S}{q}, \HT{r}{S}{q}    }
        { \HT{p \Or r}{S}{q}            }\]
\end{AuxRule}

\begin{AuxRule} CONJUNCTION \mylabel{rul:conj}
\[ \frac{ \HT{p_1}{S}{q_1}, \HT{p_2}{S}{q_2}     }
        { \HT{p_1 \A p_2}{S}{q_1 \A q_2}         } \]
\end{AuxRule}

\begin{AuxRule} $\te$-INTRODUCTION \mylabel{rul:intro}
\[ \frac{ \HT{p}{S}{q}          }
        { \HT{\te x:p}{S}{q}    }
\]
where $x \not\in var(D) \cup var(S) \cup \mathit{free}(q)$.
\end{AuxRule}

\begin{AuxRule} INVARIANCE \mylabel{rul:inv2}
\[ \frac{ \HT{r}{S}{q}           }
        { \HT{p \A r}{S}{p \A q} } 
\]
where $\mathit{free}(p) \myI (change(D) \cup change(S)) =\ES$.
\end{AuxRule}

\begin{AuxRule} SUBSTITUTION \mylabel{rul:sub}
\[ \frac{ \HT{p}{S}{q} }  
        { \HT{p[\bar{z}:=\bar{t}]}{S}{q[\bar{z}:=\bar{t}]} } 
\]
where $var(\bar{z}) \myI
          (var(D) \cup var(S)) = var(\bar{t}) \myI
          (change(D) \cup change(S))=\ES$.
\end{AuxRule}

\subsection{Axioms and proof rules for object-oriented programs}
\label{appendix-B3}

The following axioms and proof rules were introduced 
for the object-oriented programs.

\begin{Axiom} ASSIGNMENT TO INSTANCE VARIABLES \mylabel{rul:ass-inst}
\[ \HT{p[u:=t]}{u:=t}{p} \]
where $u$ is a simple or subscripted instance variable.
\end{Axiom}

\begin{Rule}  WEAKENING  \mylabel{rul:weak}
\[
\frac{\HT{p\wedge s\not=\nulll}{s.m(\bar{t})}{q}}
{\HT{p}{s.m(\bar{t})}{q}}
\]
\end{Rule}


\begin{Rule}  RECURSION I \mylabel{rul:rec1}
\[
\begin{array}{l}
\HT{p_1}{s_1.m_1(\bar{t}_1)}{q_1},\ldots,\HT{p_n}{s_n.m_n(\bar{t}_n)}{q_n} \vdash \HT{p}{S}{q},                    \\
\HT{p_1}{s_1.m_1(\bar{t}_1)}{q_1},\ldots,\HT{p_n}{s_n.m_n(\bar{t}_n)}{q_n} \vdash \\
\qquad \HT{p_i}{\block{\local \this,\bar{u}_i:=s_i,\bar{t}_i; S_i}}{q_i}, \ i \in \{1, \LL, n\} \\
[-\medskipamount]
\hrulefill                                                      \\
\HT{p}{S}{q} 
\end{array}
\]
where $m_i(\bar{u}_i) ::S_i\in D$  for $i\in\{1,\ldots,n\}$.
\end{Rule}

\begin{Rule}  RECURSION II \mylabel{rul:rec2}
\[
\begin{array}{l}
\HT{p_1}{s_1.m_1(\bar{t}_1)}{q_1},\ldots,\HT{p_n}{s_n.m_n(\bar{t}_n)}{q_n} \vdash \HT{p}{S}{q},                    \\
\HT{p_1}{s_1.m_1(\bar{t}_1)}{q_1},\ldots,\HT{p_n}{s_n.m_n(\bar{t}_n)}{q_n} \vdash \\
\qquad \HT{p_i}{\block{\local \this,\bar{u}_i:=s_i,\bar{t}_i;\  S_i}}{q_i}, \ i \in \{1, \LL, n\} \\
p_i \ra s_i\not=\nulll, \ i \in \{1, \LL, n\} \\
[-\medskipamount]
\hrulefill                                                      \\
\HT{p}{S}{q} 
\end{array}
\]
where $m_i(\bar{u}_i) ::S_i\in D$ for $i\in\{1,\ldots,n\}$.
\end{Rule}



\subsection{Proof rule for recursive programs}
\label{appendix-B4}

Finally, the following proof rule was  introduced 
for the recursive programs.


\begin{Rule}  RECURSION III \mylabel{rul:rec-proc}
\[
\begin{array}{l}
\HT{p_1}{P_1(\bar{t}_1)}{q_1},\ldots,\HT{p_n}{P_n(\bar{t}_n)}{q_n} \vdash \HT{p}{S}{q},                    \\
\HT{p_1}{P_1(\bar{t}_1)}{q_1},\ldots,\HT{p_n}{P_n(\bar{t}_n)}{q_n} \vdash \\
\qquad \HT{p_i}{\block{\local \bar{u}_i:=\bar{t}_i; S_i}}{q_i}, \ i \in \{1, \LL, n\} \\
[-\medskipamount]
\hrulefill                                                      \\
\HT{p}{S}{q} 
\end{array}
\]
where $P_i(\bar{u}_i) ::S_i\in D$.
\end{Rule}

\section{Proof systems}
\label{appendix-C}

In the following we list the proof systems used in this paper.

\subsection*{Kernel language}

\NI
PROOF SYSTEM {\it PK} for partial correctness:

\medskip

\begin{minipage}{11cm}
This system consists of the group of axioms and rules
\myref{rul:skip}--\myref{rul:fail1}, and \myref{rul:block}.
\end{minipage}
\III

\NI
PROOF SYSTEM {\it SPK} for strong partial correctness:

\medskip

\begin{minipage}{11cm}
This system consists of the group of axioms and rules
\myref{rul:skip}--\myref{rul:cons}, 
\myref{rul:fail2}, and \myref{rul:block}.
\end{minipage}

\subsection*{Object-Oriented Programs}

\NI
PROOF SYSTEM {\it PO} for partial correctness:

\medskip

\begin{minipage}{11cm}
This system is obtained by extending \emph{PK} with 
the axiom~\myref{rul:ass-inst} for assignments to instance variables,
the weakening rule~\myref{rul:weak},
and the auxiliary rules~\ref{rul:disj}--\ref{rul:sub}.
\end{minipage}
\III

\NI
PROOF SYSTEM ${\it PO}^+$ for strong partial correctness:

\medskip

\begin{minipage}{11cm}
This system is  obtained by extending {\it PO} by
the recursion I rule~\myref{rul:rec1}.
\end{minipage}
\III

\NI
PROOF SYSTEM {\it SPO} for partial correctness:

\medskip

\begin{minipage}{11cm}
This system is obtained by extending \emph{SPK} with 
the axiom~\myref{rul:ass-inst} for assignments to instance variables
and the auxiliary rules~\ref{rul:disj}--\ref{rul:sub}.
\end{minipage}
\III

\NI
PROOF SYSTEM ${\it SPO}^+$ for strong partial correctness:

\medskip

\begin{minipage}{11cm}
This system is  obtained by extending 
{\it SPR} by the recursion II rule~\myref{rul:rec2}.
\end{minipage}

\subsection*{Recursive Programs}

\NI
PROOF SYSTEM {\it PR} for partial correctness:

\medskip

\begin{minipage}{11cm}
This system is obtained by extending \emph{PK} with 
the auxiliary rules~\ref{rul:disj}--\ref{rul:sub}.
\end{minipage}
\III

\NI
PROOF SYSTEM {\it SPR} for strong partial correctness:

\medskip

\begin{minipage}{11cm}
This system is obtained by extending \emph{SPK} with 
the auxiliary rules~\ref{rul:disj}--\ref{rul:sub}.
\end{minipage}
\III

\NI
PROOF SYSTEM ${\it PR}^+$ for partial correctness:

\medskip

\begin{minipage}{11cm}
This system is  obtained by extending 
{\it PR}  by the recursion~III rule~\myref{rul:rec-proc}.
\end{minipage}
\III

\NI
PROOF SYSTEM ${\it SPR}^+$ for strong partial correctness:

\medskip

\begin{minipage}{11cm}
This system is  obtained by extending the 
{\it SPR} by the recursion~III rule~\myref{rul:rec-proc}.
\end{minipage}


\bibliographystyle{abbrv}

\bibliography{apt,abo-book,new}

\end{document}